\documentclass[11pt,twocolumn]{article}

\usepackage{graphicx}
\usepackage{algorithm,algorithmic}
\usepackage{amsmath}
\usepackage{multirow}
\usepackage{float}
\usepackage{url}
\usepackage[none]{hyphenat}
\usepackage{authblk}

\usepackage{txfonts,balance}

\usepackage{afterpage}

\newcommand\Ta{\rule{0pt}{4.0ex}}
\newcommand\Ba{\rule[-2.2ex]{0pt}{0pt}}

\begin{document}
\newcounter{cntr1}
\newcounter{cntr2}
\emergencystretch 3em

\title{GSEIM: A General-purpose Simulator with Explicit and Implicit Methods}

\author[1]{Mahesh~B.~Patil}
\author[1]{Ruchita~D.~Korgaonkar}
\author[1]{Kumar~Appaiah}
\affil[1]{Department of Electrical Engineering, Indian Institute of Technology Bombay}

\maketitle

\begin{abstract}
A new simulation package, GSEIM, for solving a set of ordinary
differential equations is presented. The organisation of the
program is illustrated with the help of a block diagram. Various
features of GSEIM are discussed. Two ways of incorporating new
elements in GSEIM, viz., as a template and as a subcircuit, are
explained by taking a specific example. Simulation examples
are described to bring out the capabilities of GSEIM.
\end{abstract}

\section{Introduction}
A wide variety of engineering applications require numerical solution
of a set of ordinary differential equations (ODEs), satisfying some
given initial conditions. This need is currently addressed by
commercial\,\cite{simulink},\cite{dymola}
as well as open-source\,\cite{xcos}
software packages.
While the numerical methods for solving ODEs are well-known
(see, e.g.,
\cite{mccalla}-%
\nocite{shampine}%
\nocite{gerald}%
\nocite{chapra}%
\nocite{burden}%
\cite{mbpvrvtr}),
different packages have different strengths and weaknesses, based
on performance, ease of use, capability of adding new library
elements, cost, user support, and legacy issues. It is the purpose
of this paper to present a new ODE solver called GSEIM (General-purpose
Simulator with Explicit and Implicit Methods) and illustrate its
working through examples.

The paper is organised as follows.
In Sec.~\ref{sec_expimp}, we briefly review the advantages and
limitations of explicit and implicit methods for solving ODEs.
We then describe, in Sec.~\ref{sec_org}, the block-level organisation of
GSEIM. A key feature of GSEIM is the flexibility it gives the user to
add new elements to the library. We describe this aspect in
Sec.~\ref{sec_templates} where we point out, using a few examples,
how computations related to explicit and implicit methods are
incorporated in the element templates.
In Sec.~\ref{sec_subc}, we look at how a subcircuit (hierarchical block)
can be added to GSEIM, using the example of an induction machine model.
One important requirement in many engineering applications is accurate
handling of abrupt changes. We describe in Sec.~\ref{sec_abrupt} how that
is implemented in GSEIM. In Sec.~\ref{sec_examples}, we present two
simulations examples to illustrate the capabilities of the new platform.
Finally, in Sec.~\ref{sec_conclusions}, we present our conclusions and
comment on future directions.

\section{Explicit and implicit methods}
\label{sec_expimp}
There are several well-known explicit and implicit methods for
solving ODEs (e.g., see \cite{shampine}). In order to illustrate
the advantage of an explicit method over an implicit method, let us
consider a single ODE of the form
\begin{equation}
\displaystyle\frac{dx}{dt} = f(t,x).
\label{eq_ode}
\end{equation}
The discretised form of this ODE using the improved Euler
method (an explicit method) is given by
\begin{equation}
x_{n+1} = x_n + \displaystyle\frac{h}{2}\,
\left[
f(t_n,x_n) +
f(t_n+h,x_n+h\,f(t_n,x_n))
\right],
\label{eq_meuler}
\end{equation}
where $x_n$ and $x_{n+1}$ correspond to the numerical solutions
at times $t_n$ and $t_{n+1}$, respectively, and
$h \,$=$\, t_{n+1}-t_n$ is the time step. Since $t_n$ and $x_n$
are known, computing $x_{n+1}$ involves only {\it{evaluation}} in this case.

Consider now the discretised form of Eq.~\ref{eq_ode} when the
backward Euler method (an implicit method) is used:
\begin{equation}
x_{n+1} = x_n + h\,f(t_{n+1},x_{n+1}).
\label{eq_beuler}
\end{equation}
Because $x_{n+1}$ is also involved on the right-hand side,
obtaining $x_{n+1}$ in this case requires {\it{solution}} of
Eq.~\ref{eq_beuler}.

For a system of ODEs, an explicit
method would still involve function evaluations, i.e., the process of
updating the system variables by evaluating functions of their past values. An
implicit method, on the other hand, would give rise to a system of
equations which has to be {\it{solved}}. If the system of equations is nonlinear, 
an iterative procedure such as the Netwon-Raphson method would be required
with the associated complication of convergence difficulties. Clearly, from
the perspective of work per time step, an explicit method would be
advantageous over an implicit method of the same order. However, implicit
methods are superior in terms of stability, as illustrated by the following
example.

\begin{figure}[!ht]
\centering
\scalebox{0.9}{\includegraphics{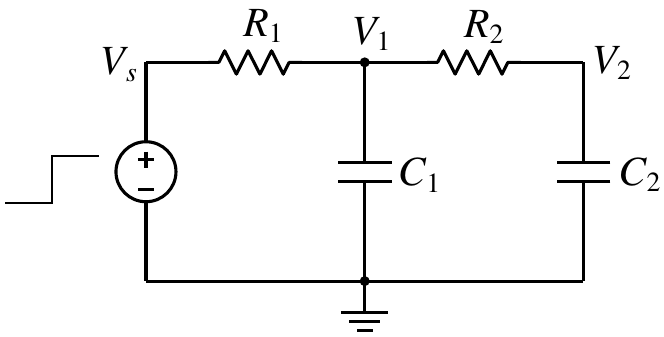}}
\vspace*{-0.2cm}
\caption{$RC$ circuit with two time constants.}
\label{fig_rc}
\end{figure}

Consider the $RC$ circuit shown in Fig.~\ref{fig_rc}.
We are interested in the variation of $V_1$ and $V_2$
when a step input voltage $V_s(t)$ is applied.
For solving the circuit equations numerically, we first
rewrite them as a system of ODEs,
\begin{equation}
 \begin{array}{rl}
  \displaystyle\frac{dV_1}{dt}
  &\hspace{-2mm}=\hspace{0mm}
  \displaystyle\frac{1}{R_1C_1}\,(V_s-V_1)-
  \displaystyle\frac{1}{R_2C_1}\,(V_1-V_2),
  \\
  \Ba \Ta
  \displaystyle\frac{dV_2}{dt}
  &\hspace{-2mm}=\hspace{0mm}
  \displaystyle\frac{1}{R_2C_2}\,(V_1-V_2).
 \end{array}
 \label{eq_rc}
\end{equation}

We expect $V_1$ and $V_2$ to start changing as the input step
is applied and eventually settle down to their steady-state
values. For this problem, rather than using constant time steps
for the entire interval of interest, it is far more efficient to use
small time steps when the variations are rapid and large time steps
when they are slow.
We consider two methods which employ such adaptive time step
computation: 
(a)~the Runge-Kutta-Fehlberg (RKF45) method, an explicit
method which employs Runge-Kutta methods of order 4 and 5
in each time step\,\cite{burden}, and
(b)~the TR-BDF2 method, an implicit method, which employs the
trapezoidal (TR) and backward difference formula (BDF2) methods
in each time step\,\cite{bank}.

In each of these methods, an estimate of the local truncation
error (LTE) is obtained in each time step. If the LTE is small,
the current time step is accepted, and the next time step is
allowed to be larger. If the LTE is larger than a specified
value, the current time step is rejected, and a smaller time
step is tried. As the circuit appraches steady state, the LTE
tends to zero, allowing the algorithm to take larger time steps,
limited eventually only by an upper limit set by the user.

\begin{figure}[!ht]
\centering
\hspace*{-0.0cm}{\includegraphics[width=1.0\columnwidth]{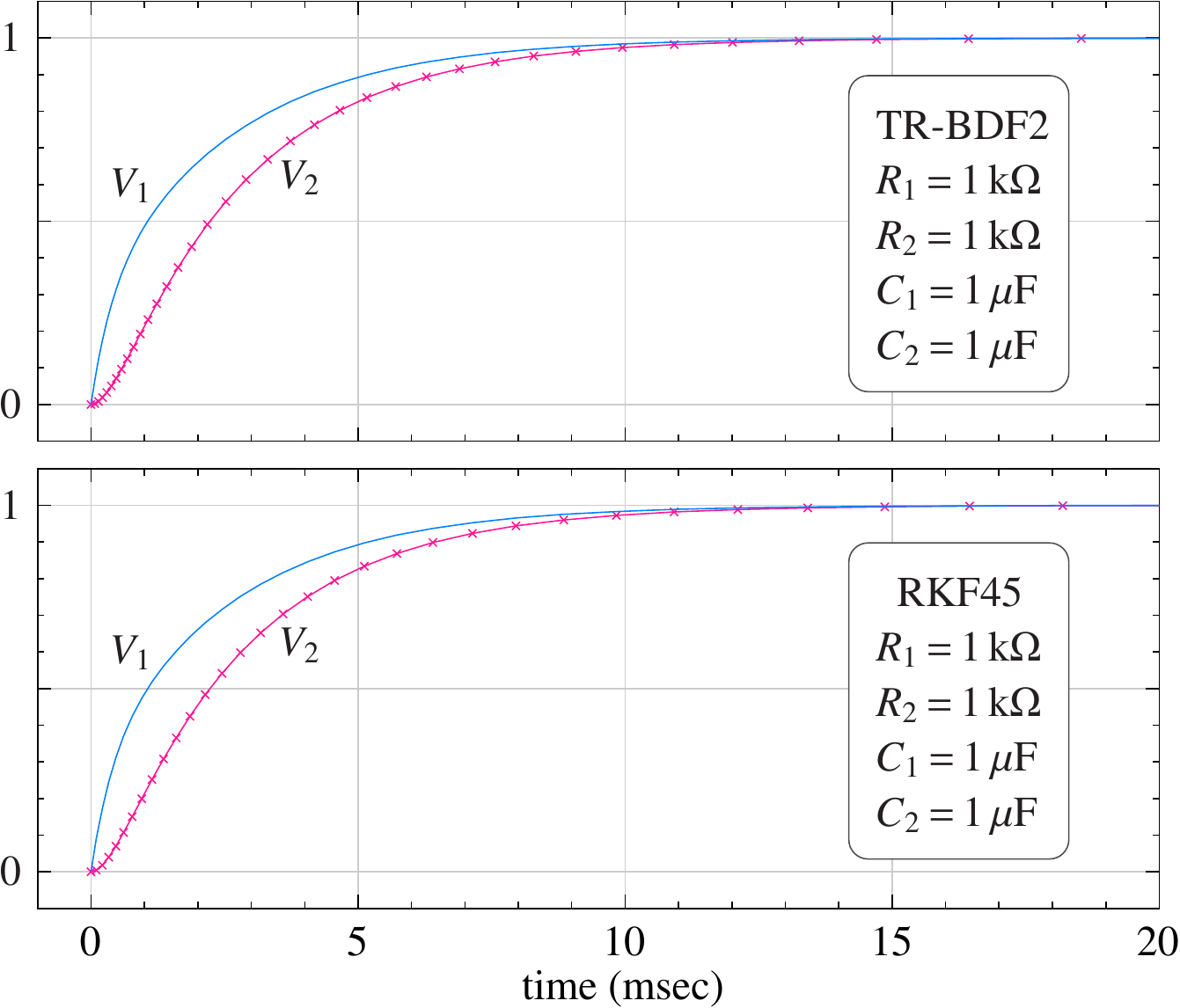}}
\vspace*{-0.2cm}
\caption{Numerical solution of Eq.~\ref{eq_rc} using the
TR-BDF2 and RKF45 methods. The parameter values are
$R_1 \,$=$\, R_2 \,$=$\, 1$\,k$\Omega$,
$C_1 \,$=$\, C_2 \,$=$\, 1\,\mu$F.
Crosses show the simulator time points.}
\label{fig_rkf1}
\end{figure}

Fig.~\ref{fig_rkf1} shows the results for
$R_1 \,$=$\, R_2 \,$=$\, 1$\,k$\Omega$,
$C_1 \,$=$\, C_2 \,$=$\, 1\,\mu$F.
Both RKF45 and TR-BDF2 methods perform as expected. As the
circuit approaches steady state, they make the time steps
larger, leading to fewer time steps overall and therefore
faster simulation.

\begin{figure}[!ht]
\centering
\hspace*{-0.0cm}{\includegraphics[width=1.0\columnwidth]{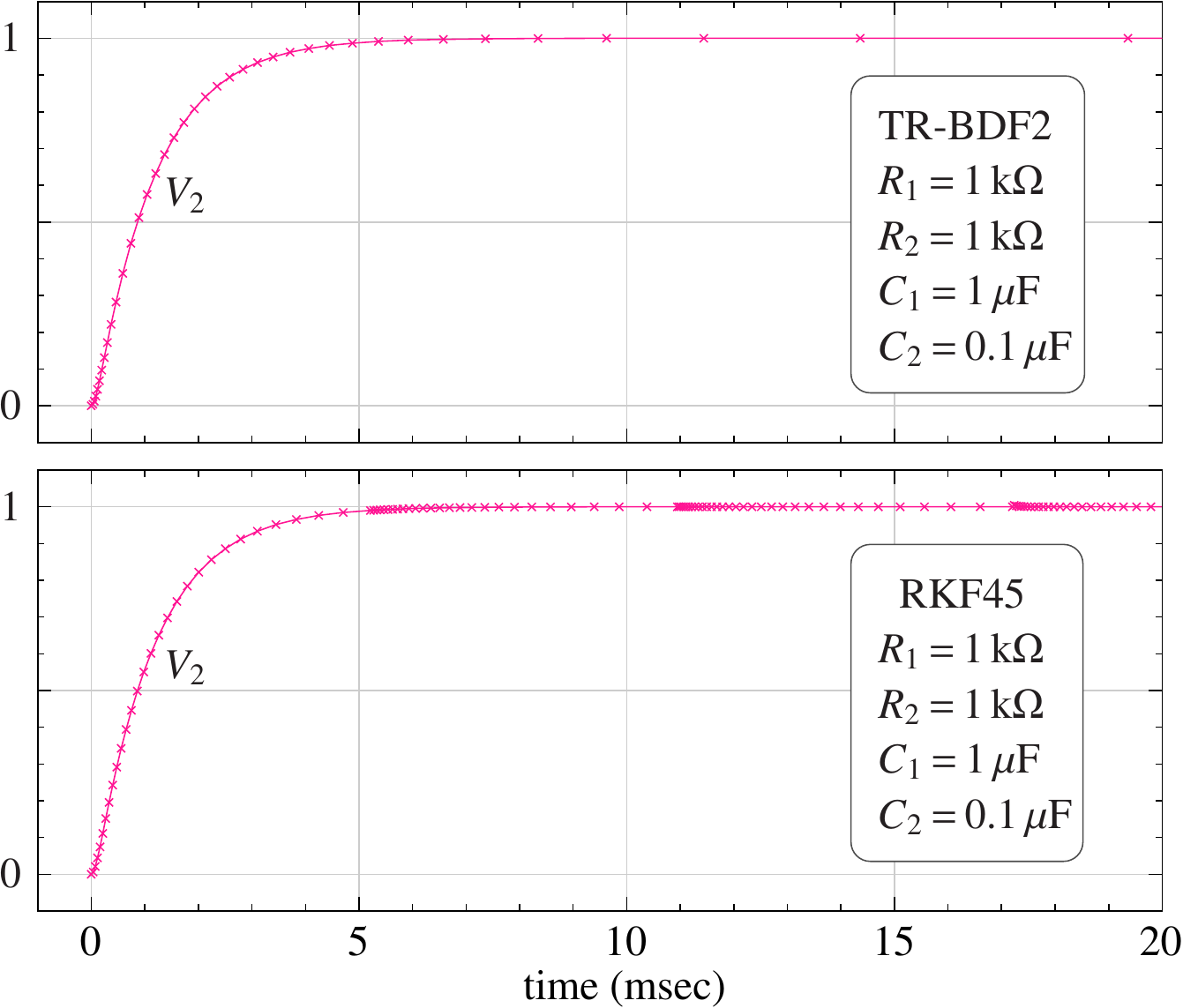}}
\vspace*{-0.2cm}
\caption{Numerical solution of Eq.~\ref{eq_rc} using the
TR-BDF2 and RKF45 methods. The parameter values are
$R_1 \,$=$\, R_2 \,$=$\, 1$\,k$\Omega$,
$C_1 \,$=$\, 1\,\mu$F,
$C_2 \,$=$\, 0.1\,\mu$F.
Crosses show the simulator time points.}
\label{fig_rkf2}
\end{figure}

When $C_2$ is changed from $1\,\mu$F to $0.1\,\mu$F, the two
methods show very different behaviour (see Fig.~\ref{fig_rkf2}).
The TR-BDF2 method
continues to increase the time step as the circuit approaches
steady state. The RKF45 method does increase the time step up to
a certain point, but at
$t \,$$\approx$$\, 5.1$\,msec, it forces a small time step. After
that, it once again starts increasing the time step, but only up
to $t \,$$\approx$$\, 10.8$\,msec, and so on.
This behaviour is related to stability of the RKF45 algorithm.
The explicit Runge-Kutta methods employed in the RKF45 algorithm
are conditionally stable. They require the time step to be
smaller than a certain multiple of the smallest time constant
in the system\,\cite{mbpvrvtr}. With
$R_1 \,$=$\, R_2 \,$=$\, 1$\,k$\Omega$,
$C_1 \,$=$\, C_2 \,$=$\, 1\,\mu$F, the time constant are
$\tau _1 \,$=$\, 2.6$\,msec and
$\tau _2 \,$=$\, 0.38$\,msec.
With $C_2 \,$=$\, 0.1\,\mu$F, the time constants are
$\tau _1 \,$=$\, 1.1$\,msec and
$\tau _2 \,$=$\, 0.09$\,msec, and the largest time step allowed
by the RKF45 algorithm is correspondingly reduced. For the
TR-BDF2 method, which is A-stable, there is no such restriction,
and therefore it allows large time steps as steady state is
approached, thus reducing the computation time significantly.

From the above discussion, it is clear that,
from the efficiency perspective,
the choice of the method (explicit or implicit)
would depend on the problem being solved. For this reason,
GSEIM incorporates explicit as well as implicit methods.

\section{GSEIM organisation}
\label{sec_org}
\begin{figure*}[!ht]
\centering
\scalebox{0.8}{\includegraphics{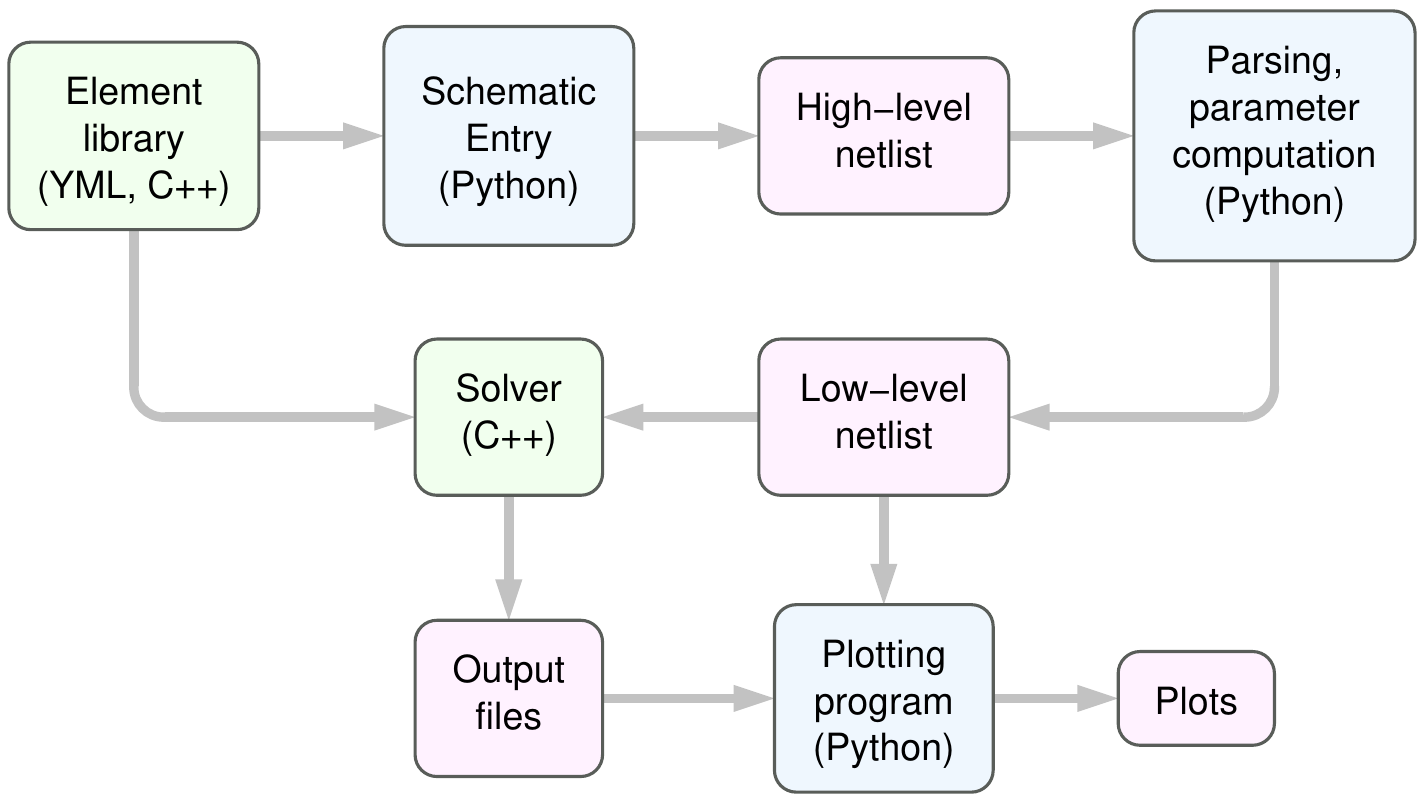}}
\vspace*{-0.2cm}
\caption{Block diagram of the GSEIM program.}
\label{fig_block}
\end{figure*}

The block diagram of the GSEIM program is shown in
Fig.~\ref{fig_block}.  The schematic entry block (GUI)
is adapted from the GNURadio package\,\cite{gnr}.
It enables the user to prepare a schematic diagram
of the system of interest and produces a high-level
netlist. A parser program takes the high-level netlist
as input, performs parsing of node names and computation
of element parameters (if required). As its output, the
parser program produces a low-level netlist.
The solver takes the low-level netlist as its input,
and using information from the library, it prepares and
solves the set of ODEs corresponding to the user's
system, creating output files requested by the user.
Finally, the plotting program reads the output files and
displays the plots in an interactive manner.

GSEIM has been designed to completely decouple the element
library from the solver, which makes it possible for the user
to add new elements to the library (if required). For each
element (say, {\tt{xyz}}), the library contains two files:
(a)~{\tt{xyz.xbe}} which contains information about the
variable names, parameter names and values, and equations
related to that element, and (b)~{\tt{xyz.yml}} which specifies
how the element would appear in the schematic entry GUI. A
detailed description of these files would be presented in the
GSEIM manual. In Sec.~\ref{sec_templates}, we will take a brief
look at the {\tt{xbe}} files for a few elements.

\begin{figure*}[!ht]
\centering
\hspace*{-0.1cm}{\includegraphics[width=1.0\textwidth]{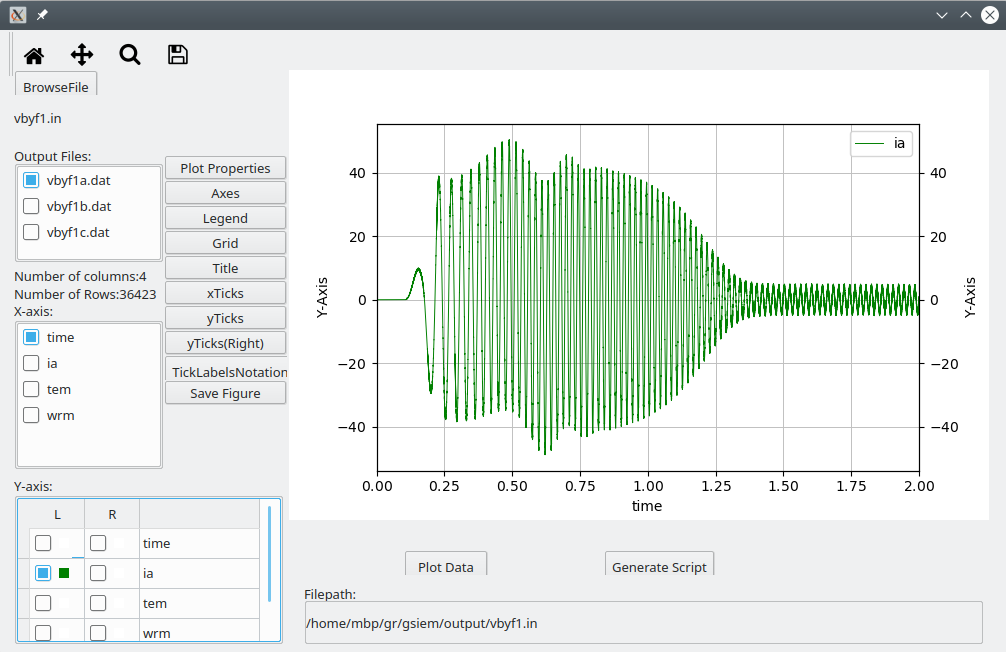}}
\caption{Snapshot of the plotting GUI provided as part of the GSEIM package.}
\label{fig_plotgui}
\end{figure*}

The salient features of the GSEIM package can be described as follows.
\begin{list}{(\alph{cntr1})}{\usecounter{cntr1}}
 \item
  The solver, which handles the most intensive computation, viz.,
  numerical solution of the ODEs, is written in C++ because of
  its high performance.
 \item
  For all other purposes, viz., schematic capture, parsing, parameter
  computation, and plotting, python is used because of the flexibility
  and ease of programming it offers.
 \item
  Output parameters, which determine what data gets stored in the output
  files during simulation, are specified without having to add extra
  elements~-- such as the ``scope" in Simulink\,\cite{simulink} and
  Xcos\,\cite{xcos})~-- to the schematic diagram. This helps in avoiding clutter.
 \item
  Subcircuits (hierarchical blocks) can be used for simplifying the
  schematic.
 \item
  Explicit as well as implicit numerical methods are incorporated in the
  solver. Currently, the following methods are made available:
  \begin{list}{(\roman{cntr2})}{\usecounter{cntr2}}
   \item
    explicit (fixed time step): improved Euler, Heun, Runge-Kutta (4$^{\mathrm{th}}$ order)
   \item
    explicit (auto time step):
    RKF45, Bogacki and Shampine (2,3)
   \item
    implicit (fixed time step):
    backward Euler, trapezoidal
   \item
    implicit (auto time step):
    backward Euler-auto, trapezoidal-auto, TR-BDF2
  \end{list}
  The backward Euler-auto and trapezoidal-auto methods are used
  only for nonlinear problems. In these methods, the time step is
  adjusted depending on the number of Newton-Raphson iterations
  required at a given time point.
 \item
  GSEIM provides a GUI for plotting the variables specified by the
  user. The plotting GUI, shown in Fig.~\ref{fig_plotgui}, allows the user
  to select the output file of interest, the $x$-axis variable
  (typically time), and $y$-axis variable(s) to be included
  in the plot. It also gives the user control over plot attributes
  such as line colour, line width, and symbol type, with appropriate
  values specified by default. Using this information, the plotting
  GUI displays the plot requested by the user.
  It also produces python code associated
  with the plot. If required, the user can edit this code in order to make the plot
  more suitable for a report or a presentation. This way,
  the user can benefit from a wide range of plotting capabilities
  offered by python.
\end{list}

\section{Library element templates}
\label{sec_templates}
A very important feature of GSEIM is that it allows the user to
add new functionality in the form of library elements, by
writing suitable ``templates."
In this section, we look at the syntax of element templates
with the help of some examples. We start with a few remarks.
\begin{list}{(\alph{cntr2})}{\usecounter{cntr2}}
 \item
  An element template has three types of variables in general:
  input, output, and auxiliary. Only the input and output variables
  are made available in the schematic capture GUI for connection to
  other elements.
 \item
  Two types of elements are allowed:
  \begin{list}{(\roman{cntr2})}{\usecounter{cntr2}}
   \item
    {\tt{evaluate}} type in which the element equations are of the form
    $y \,$=$\, f(x_1,x_2,..)$
    where $y$ is an output and $x_1$, $x_2$, etc. are inputs. These
    elements do not involve time derivatives.
   \item
    {\tt{integrate}} type with equations of the form
    $\displaystyle\frac{dy}{dt} \,$=$\,f(x_1,x_2,..)$ where $y$ is an
    output or auxiliary variable, and
    $x_1$, $x_2$, etc. can be input, output, or auxiliary variables.
  \end{list}
\end{list}
\subsection{Adder}
\label{sec_adder}
\begin{figure}[!ht]
\centering
\scalebox{0.8}{\includegraphics{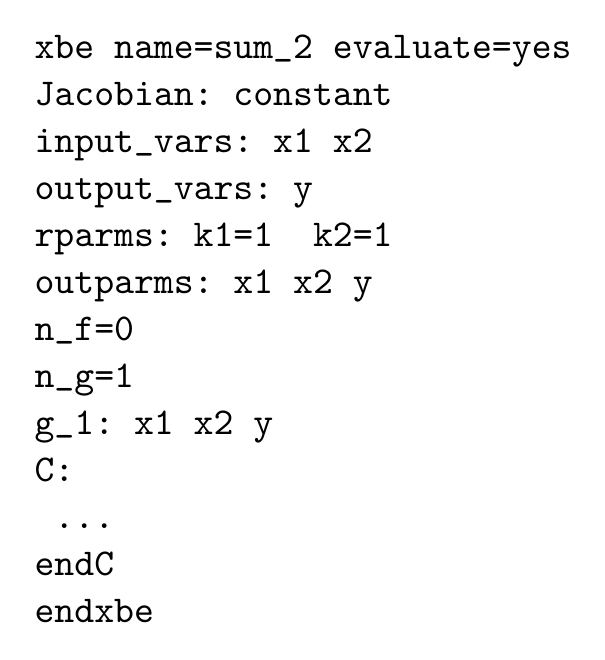}}
\vspace*{-0.2cm}
\caption{{\tt{sum\_2.xbe}} template (partial).}
\label{fig_sum2a}
\end{figure}

As out first example, we consider the {\tt{sum\_2}} element which gives
$y \,$=$\, k_1x_1 + k_2x_2$, where $x_1$ and $x_2$ are input variables,
$y$ is the output variable, and $k_1$, $k_2$ are real parameters.
This is an {\tt{evaluate}} type element, i.e., its output can be written
as a function of its inputs, and it does not involve time derivatives.
Fig.~\ref{fig_sum2a} shows the overall structure of {\tt{sum\_2.xbe}}.
The following features may be noted.
\begin{list}{(\alph{cntr2})}{\usecounter{cntr2}}
 \item
  The element name is specified by the keyword {\tt{name}}.
 \item
  The assignment {\tt{evaluate=yes}} specifies the element
  type.
 \item
  The assignment {\tt{Jacobian:~constant}} indicates that,
  when the element equation
  $y - k_1x_1 - k_2x_2 = 0$
  is differentiated with respect to the variables involved
  in the equation, we get constants.
 \item
  The lines {\tt{input\_vars}} and {\tt{output\_vars}} specify
  the input and output variables of the element, respectively.
 \item
  The names and default values of the real parameters are given
  by the {\tt{rparms}} statement. (GSEIM also allows integer and
  string parameters; they are not used in {\tt{sum\_2}}.
 \item
  The {\tt{outparms}} statement specifies the names of output
  parameters which will be made available by this template
  for saving to the user's output files
  during simulation (if requested by the user).
 \item
  The {\tt{n\_f}} and {\tt{n\_g}} statements specify the number
  of $f$ and $g$ functions for this element. (This aspect will be
  described in detail in the GSEIM manual.)
 \item
  The {\tt{g\_1}} statement indicates the variables involved in
  the function $g_1$.
 \item
  The C++ part of the template, to be described separately, appears
  between the {\tt{C}} and {\tt{endC}} statements.
\end{list}

\begin{figure}[!ht]
\centering
\scalebox{0.8}{\includegraphics{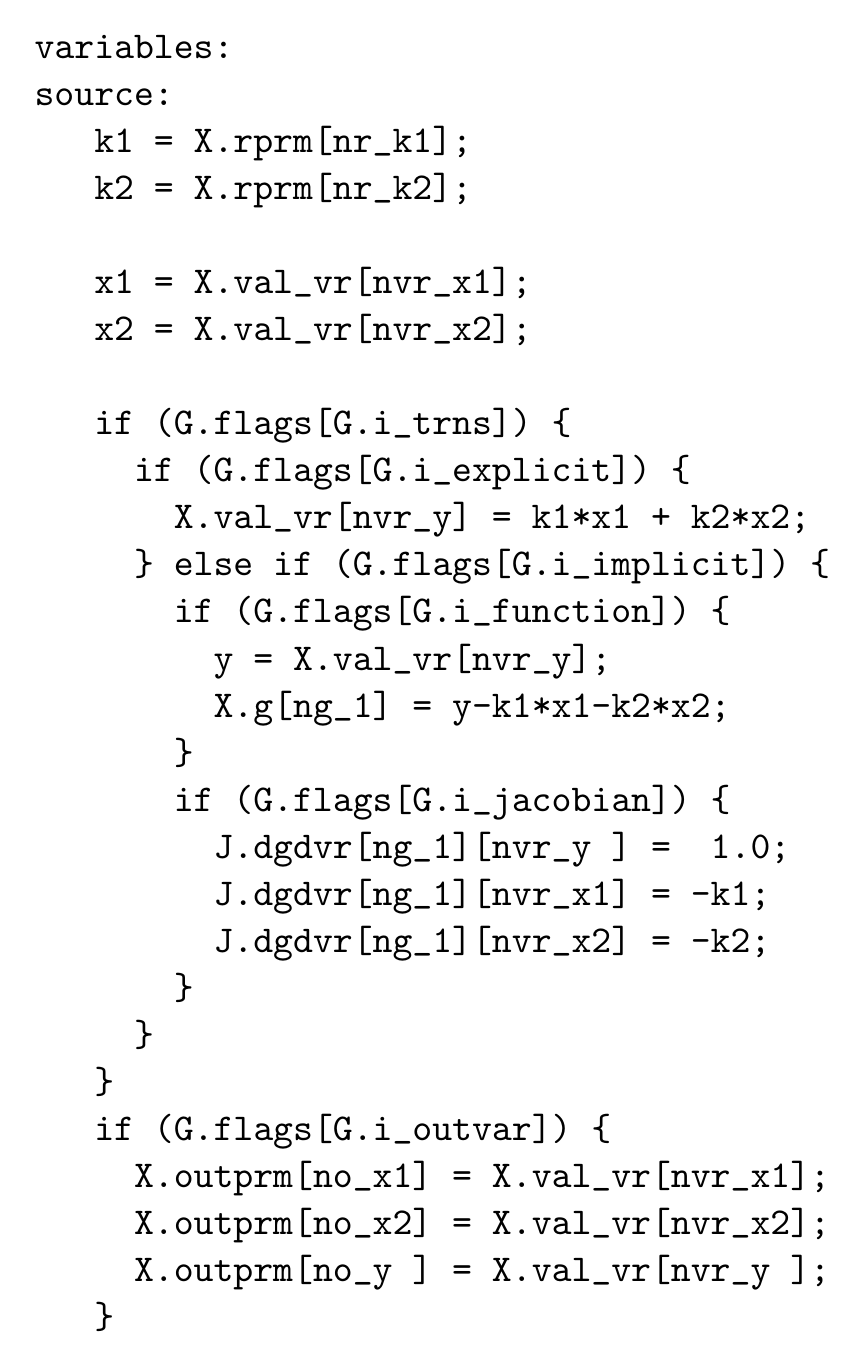}}
\vspace*{-0.2cm}
\caption{C++ part of the {\tt{sum\_2.xbe}} template (partial).}
\label{fig_sum2b}
\end{figure}

Before we look at the C++ part of {\tt{sum\_2.xbe}}, let us see where
it fits in the overall scheme. The GSEIM library preprocessor
embeds the C++ part of each element template in the C++ function
corresponding to that element. This function receives objects {\tt{X}}
and {\tt{G}} from the GSEIM main program and is expected to compute
various quantities such function values, output parameters, etc.
The object {\tt{G}} is a global object and is used to pass information
about the current time point, type of method being used (implicit or
explicit), etc. It also conveys to the element routine, through the
{\tt{flags}} array, what computation the main program is expecting
from the element routine in the present call. The object {\tt{X}} is
specific to the element being treated, and it contains variables and
parameter values related to that element.
With this background, we can make the following points about the C++
part of {\tt{sum\_2.xbe}}, as shown in Fig.~\ref{fig_sum2b}.
\begin{list}{(\alph{cntr2})}{\usecounter{cntr2}}
 \item
  If an explicit method is being used, the template only needs to
  evaluate {\tt{y}} in terms of {\tt{x1}} and {\tt{x2}}.
 \item
  If an implicit method is being used, the template needs to supply
  information about the equation it satisfies which in this case is
\begin{equation}
 g_1 \equiv y-k_1x_1-k_2x_2 = 0.
 \label{eq_sum2}
\end{equation}
  If the program is requesting the function value, $g_1(x_1,x_2,y)$ is
  evaluated; if it is requesting the derivatives, then
  $\displaystyle\frac{\partial g_1}{\partial x_1}$,
  $\displaystyle\frac{\partial g_1}{\partial x_2}$,
  $\displaystyle\frac{\partial g_1}{\partial y}$
  are evaluated.
 \item
  If the program is requesting assignment of output parameters, the
  parameters listed in the {\tt{outparms}} statement (see Fig.~\ref{fig_sum2a})
  are assigned.
\end{list}
\subsection{Integrator}
\label{sec_integrator}
Next, we consider an element of type {\tt{integrate}}, viz., the
integrator which satisfies
$y \,$=$\, k\,\int x\,dt$ where $x$ and $y$ are the input and
output variables, respectively, and $k$ is a real parameter.
Since GSEIM expects the equations to be written
in the general form
$\displaystyle\frac{dy}{dt} \,$=$\,f(x_1,x_2,..)$,
we rewrite the integrator equation as
$\displaystyle\frac{dy}{dt} \,$=$\,k\,x$.
For {\tt{integrator}} type elements, we also need to specify
the initial or ``start-up" value of the state variable(s).
For the integrator, we will denote that by $y_0$.
\begin{figure}[!ht]
\centering
\scalebox{0.8}{\includegraphics{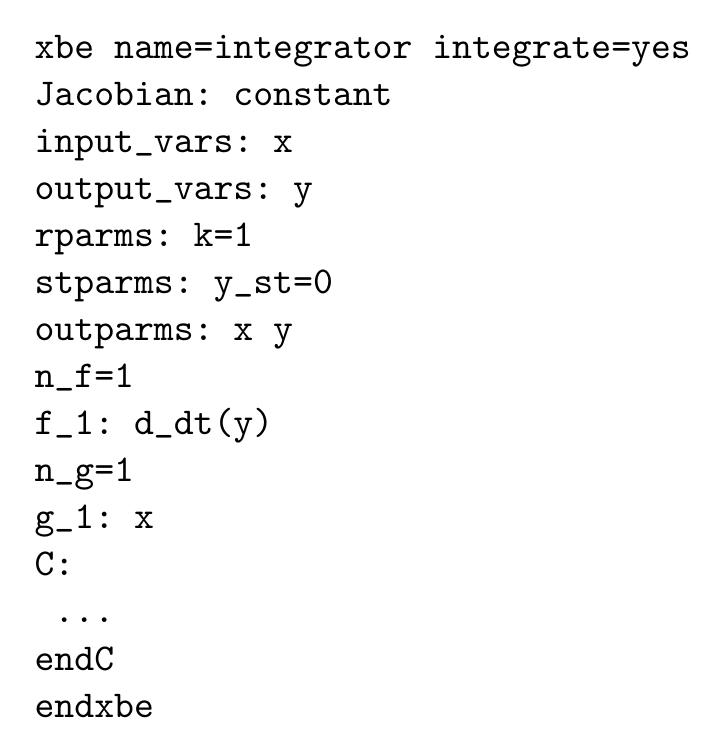}}
\vspace*{-0.2cm}
\caption{{\tt{integrator.xbe}} template (partial).}
\label{fig_integrator2a}
\end{figure}
\begin{figure}[!ht]
\centering
\scalebox{0.8}{\includegraphics{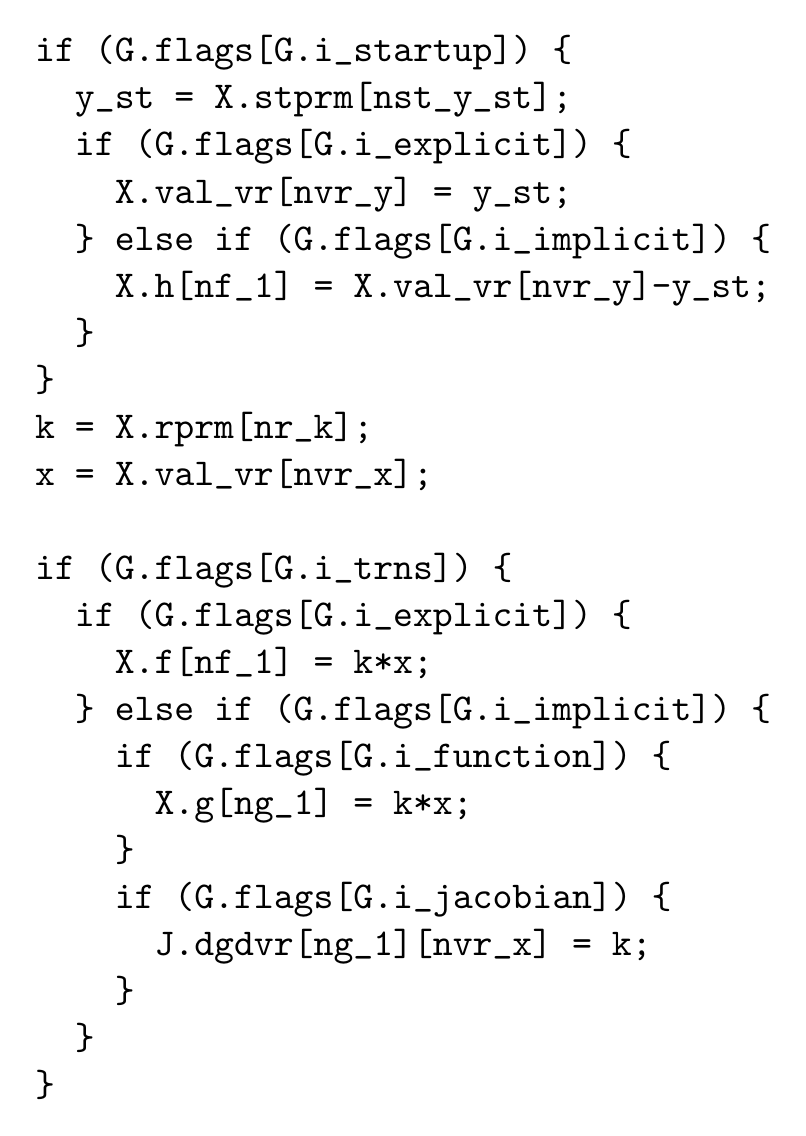}}
\vspace*{-0.2cm}
\caption{C++ part of the {\tt{integrator.xbe}} template (partial).}
\label{fig_integrator2b}
\end{figure}

The integrator template without the C++ part is shown in
Fig.~\ref{fig_integrator2a}. The C++ part is shown
separately in Fig.~\ref{fig_integrator2b}. The start-up
parameter {\tt{y\_st}} corresponds to $y_0$ mentioned above.
The fact that time derivative of {\tt{y}} is involved in the
element equation is indicated by the {\tt{f\_1}} statement.

In the C++ part of the template (Fig.~\ref{fig_integrator2b}),
we have different sections for start-up and transient simulation.
In the start-up section, the equation $y \,$=$\, y_0$ is handled.
In the transient section, if the method is explicit, only the
function $f_1 \,$=$\, k\,x$ is evaluated; if it is implicit,
the function $g_1 \,$=$\, k\,x$ as well as its derivative
$\displaystyle\frac{\partial g_1}{\partial x}$ are computed.

\subsection{Induction motor}
\label{sec_indmc}
We now look at a more complex element of type {\tt{integrate}},
viz., {\tt{indmc1.xbe}}, which implements the induction machine model
given by,
\begin{eqnarray}
\displaystyle\frac{d{\psi}_{ds}}{dt}& = &v_{ds}-r_si_{ds},
\label{eq_indmc_1} \\
\displaystyle\frac{d{\psi}_{qs}}{dt}& = &v_{qs}-r_si_{qs},
\label{eq_indmc_2} \\
\displaystyle\frac{d{\psi}_{dr}}{dt}& = &-\,\frac{P}{2}\,\omega_ {rm}\psi _{qr}-r_ri_{dr},
\label{eq_indmc_3} \\
\displaystyle\frac{d{\psi}_{qr}}{dt}& = & \frac{P}{2}\,\omega_ {rm}\psi _{dr}-r_ri_{qr},
\label{eq_indmc_4} \\
\displaystyle\frac{d{\omega}_{rm}}{dt}& = &\frac{1}{J}\,(T_{em}-T_L),
\label{eq_indmc_5}
\end{eqnarray}
\noindent
where
\begin{eqnarray}
i_{ds}& = &\frac{L_r}{L_m L_e}\,\psi _{ds} - \frac{1}{L_e}\,\psi _{dr},
\label{eq_indmc1_1} \\
i_{qs}& = &\frac{L_r}{L_m L_e}\,\psi _{qs} - \frac{1}{L_e}\,\psi _{qr},
\label{eq_indmc1_2} \\
i_{dr}& = &\frac{1}{L_m}\,\psi _{ds} - \left(\frac{L_{ls}}{L_m}+1\right)~i _{ds},
\label{eq_indmc1_3} \\
i_{qr}& = &\frac{1}{L_m}\,\psi _{qs} - \left(\frac{L_{ls}}{L_m}+1\right)~i _{qs},
\label{eq_indmc1_4} \\
T_{em}& = &\frac{3}{4}\, P L_m\,(i_{qs}i_{dr} - i_{ds}i_{qr}),
\label{eq_indmc1_5}
\end{eqnarray}
\noindent
with $L_e \,$=$\, \displaystyle\frac{L_sL_r}{L_m}-L_m$,
$L_s \,$=$\, L_{ls}+L_m$,
$L_r \,$=$\, L_{lr}+L_m$.

\begin{figure}[!ht]
\centering
\scalebox{0.8}{\includegraphics{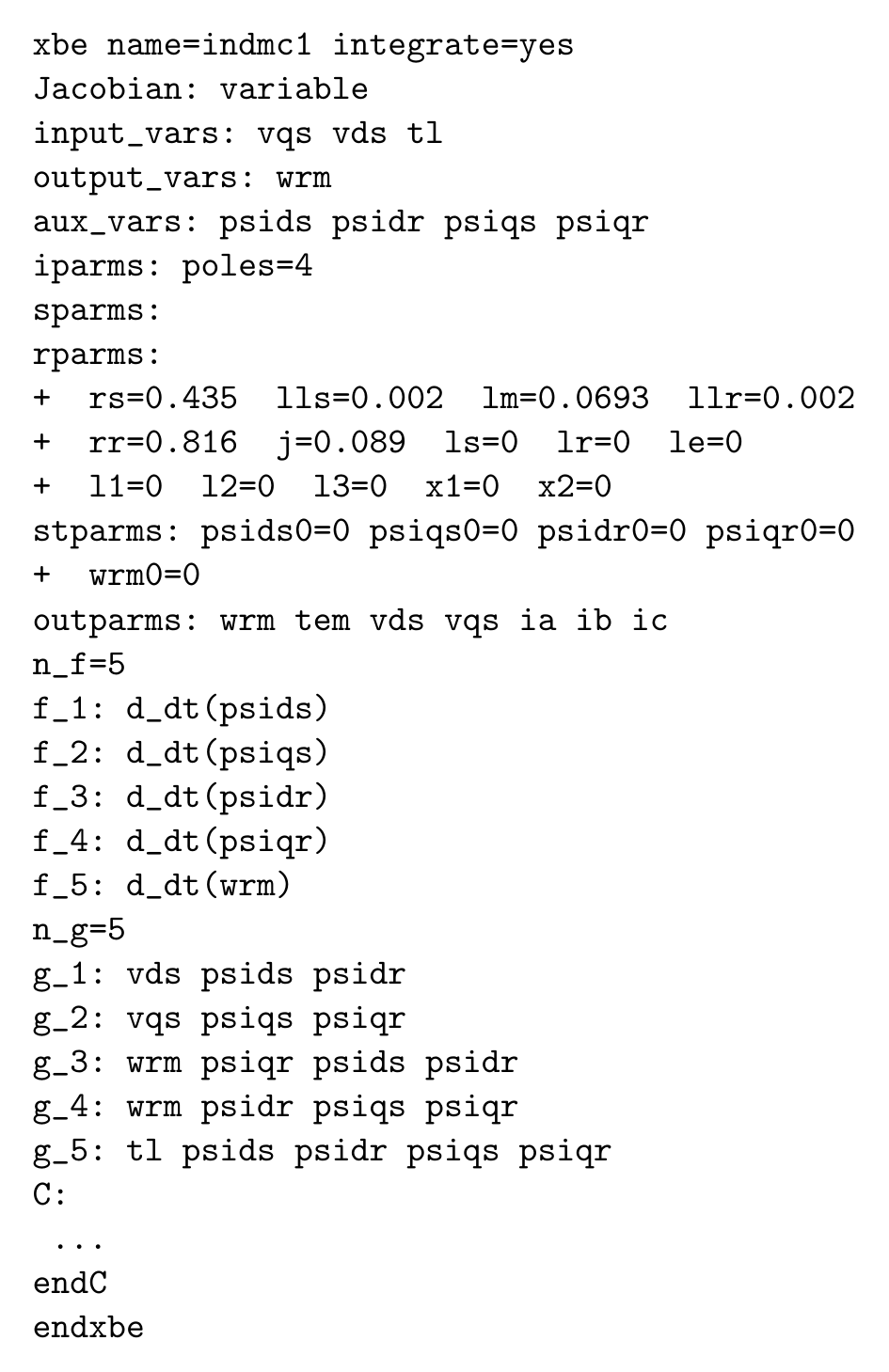}}
\vspace*{-0.2cm}
\caption{{\tt{indmc1.xbe}} template (partial).}
\label{fig_indmc1a}
\end{figure}

\begin{figure}[!ht]
\centering
\scalebox{0.8}{\includegraphics{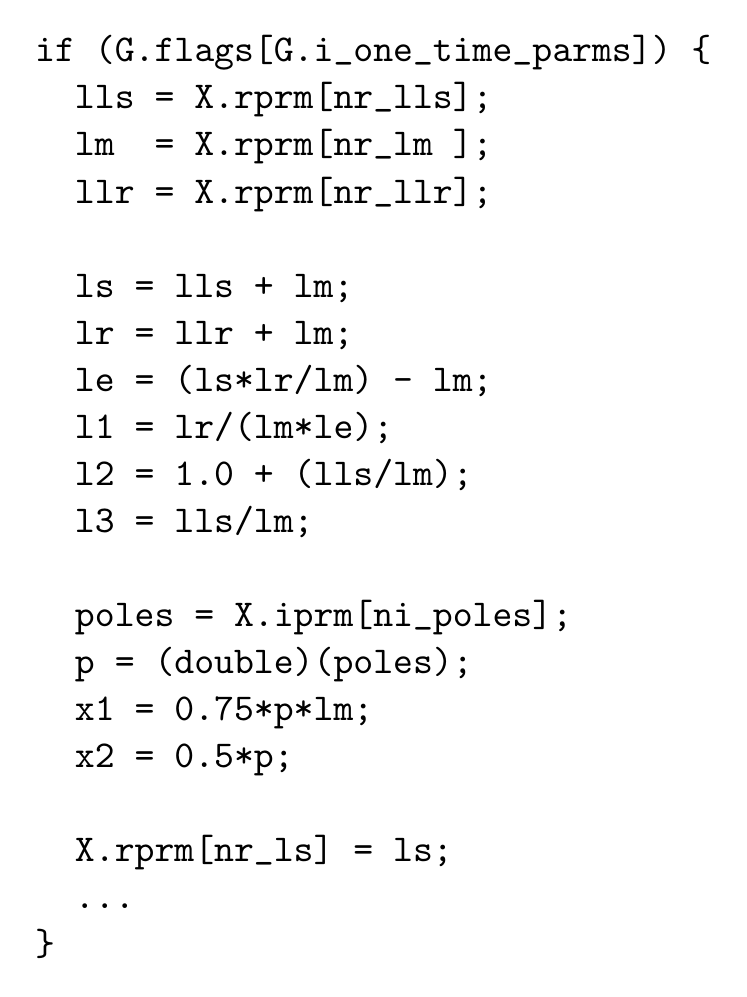}}
\vspace*{-0.2cm}
\caption{One-time parameter section of {\tt{indmc1.xbe}}.}
\label{fig_indmc1b}
\end{figure}

\begin{figure}[!ht]
\centering
\scalebox{0.8}{\includegraphics{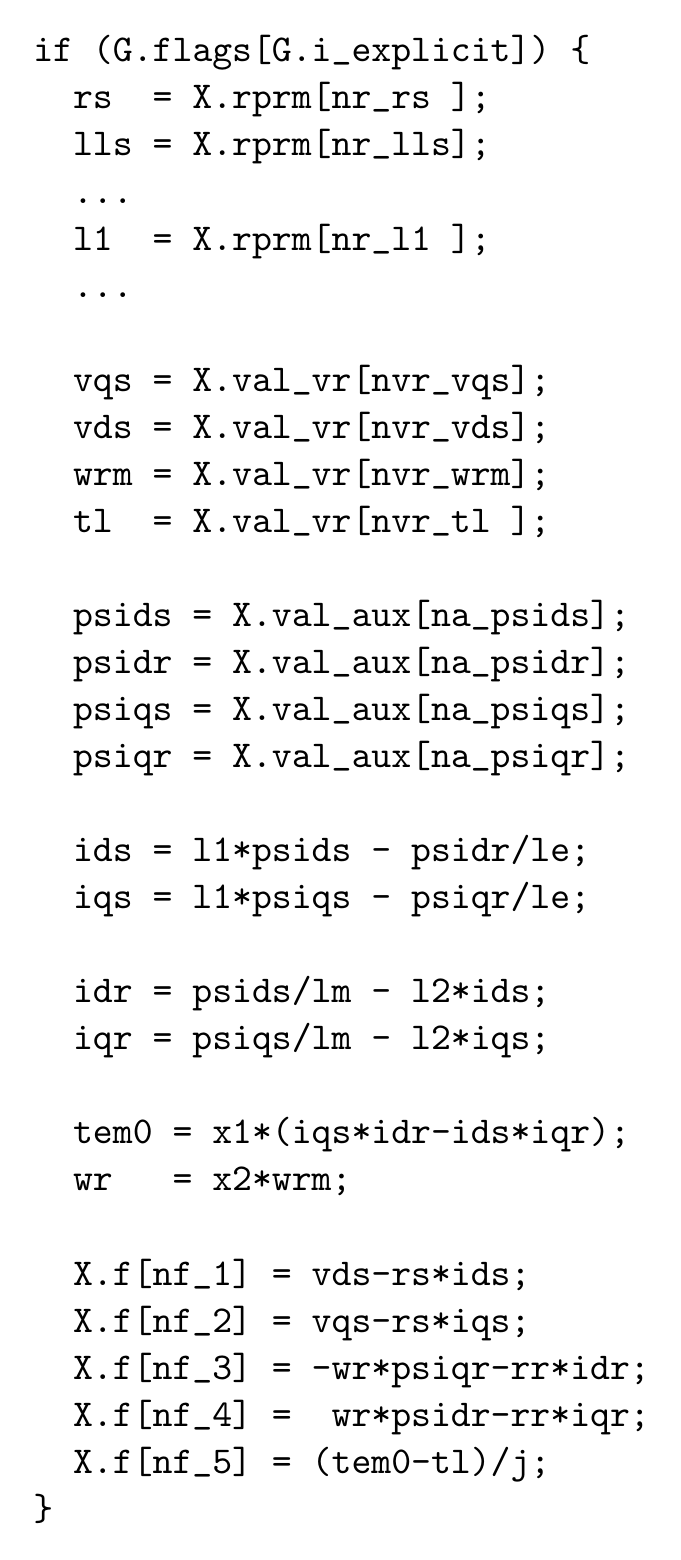}}
\vspace*{-0.2cm}
\caption{Function evaluation in {\tt{indmc1.xbe}} for explicit methods.}
\label{fig_indmc1c}
\end{figure}

\begin{figure}[!ht]
\centering
\scalebox{0.8}{\includegraphics{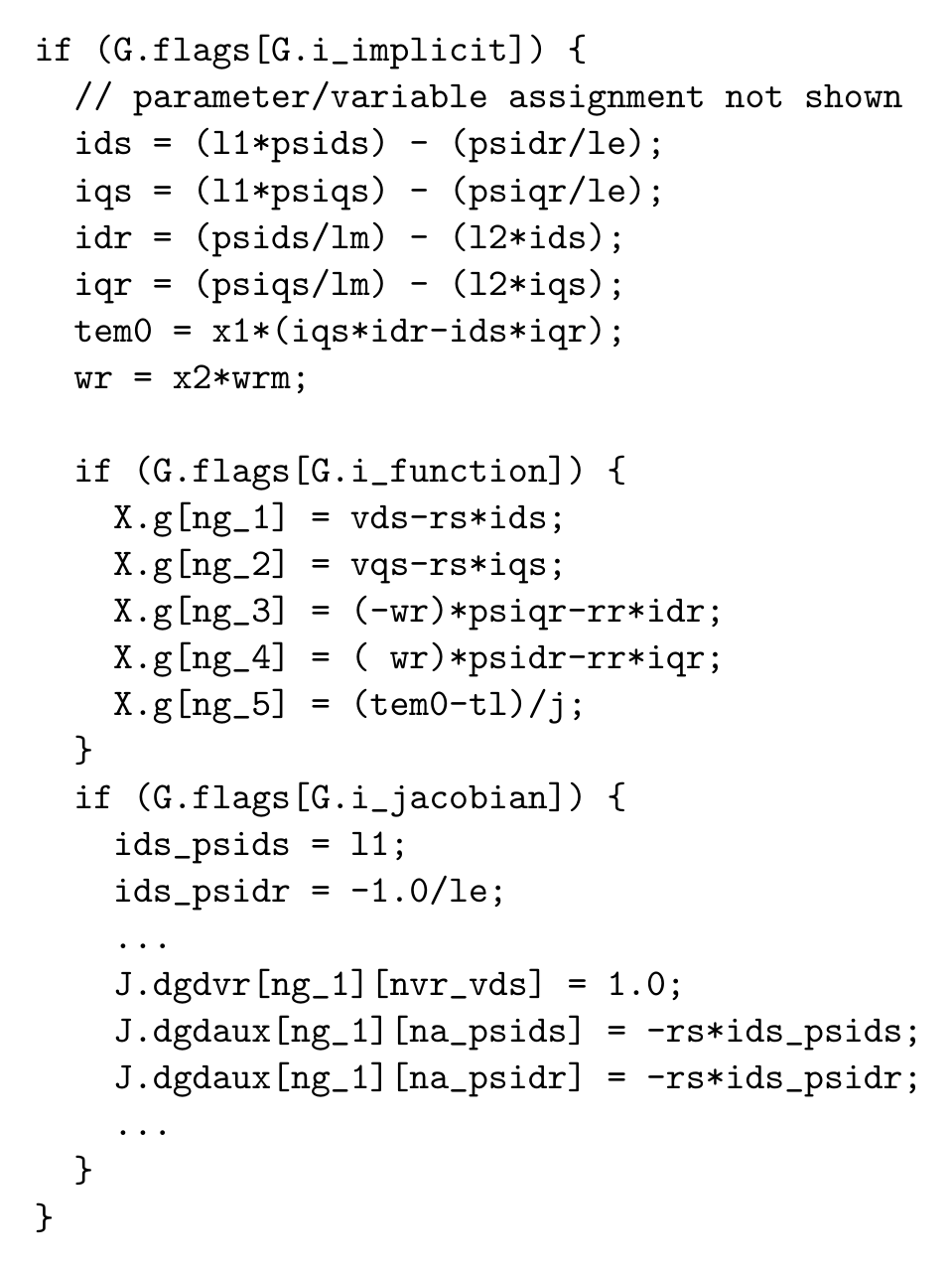}}
\vspace*{-0.2cm}
\caption{Function evaluation in {\tt{indmc1.xbe}} for implicit methods.}
\label{fig_indmc1d}
\end{figure}

Figs.~\ref{fig_indmc1a}-\ref{fig_indmc1d} show the various sections
of the {\tt{indmc1}} template. The input variables are
$v_{qs}$, $v_{ds}$, $T_L$, and the output variable is $\omega _{rm}$.
In addition, it has internal (auxiliary) variables
$\psi _{ds}$,
$\psi _{dr}$,
$\psi _{qs}$,
$\psi _{qr}$
which are involved in the model equations.
The statements {\tt{f\_1}}, {\tt{f\_2}}, etc. in Fig.~\ref{fig_indmc1a}
are used to inform the simulator which derivative is involved in that equation.
The statements {\tt{g\_1}}, {\tt{g\_2}}, etc.
are used to indicate which variables are involved in the right-hand
side of the corresponding equation.

In the induction machine equations (Eqs.~\ref{eq_indmc_1}-\ref{eq_indmc1_5}),
there are some ``one-time" calculations, e.g., calculation of $L_e$,
which are not required to be performed in every time step. GSEIM provides
a flag for this purpose, as seen in Fig.~\ref{fig_indmc1b}. When this flag is
set by the main program, the template computes $L_e$ and other one-time
parameters, and saves them in the {\tt{X.rprm}} vector. Subsequently, these
parameters need not be computed again.

The function assignment sections of {\tt{indmc1.xbe}} are shown separately in 
Figs.~\ref{fig_indmc1c} and \ref{fig_indmc1d} for explicit and implicit
methods, respectively.
In the explicit case (Fig.~\ref{fig_indmc1c}),
the function $f_1$ (i.e.,
{\tt{f[nf\_1]}}) is computed as per the right-hand side of Eq.~\ref{eq_indmc_1},
and so on.
In the implicit case (Fig.~\ref{fig_indmc1d}),
the function $g_1$ is simlarly computed. However, in this case,
the derivatives of $g_1$ with respect to each of the variables involved
in this equation also need to be computed.

As seen from the above example, writing a new template, particularly the
Jacobian assignment part, requires some systematic effort. For this reason,
some simulation packages allow the use of hardware description languages
such as Openmodelica\,\cite{openmodelica} and Verilog-AMS\,\cite{verilogams},
which require from the user only functions in symbolic form. The derivatives
are then computed internally by the simulator.

While the ``low-level" approach used in GSEIM demands more effort from the
user in developing new element templates, it does offer more flexibility~--
the user is not constrained by the limitations of
a high-level language. Furthermore, library development is a one-time
activity; once an element template is developed and tested, no further
coding is required. We believe therefore that our low-level approach
has significant practical relevance.


\section{Subcircuits}
\label{sec_subc}
In many situations, the system of interest is hierarchical in nature,
and building it in a modular fashion is easier or more convenient than
assembling all the basic blocks at one level. Like simulation packages
such as SPICE\,\cite{ngspice}, Simulink\,\cite{simulink},
Dymola\,\cite{dymola}, GSEIM also allows hierarchical system building.
In this section, we consider the induction machine model of
Sec.~\ref{sec_templates} and describe how it can be implemented as a
``subcircuit" (a hierarchical block) rather than writing an element
template. For this purpose, we rewrite
Eqs.~\ref{eq_indmc_1}-\ref{eq_indmc1_5} such that each of them can be
implemented using basic blocks such as adder, multiplier, integrator, etc.

\begin{figure*}[!ht]
\centering
\hspace*{-0.1cm}{\includegraphics[width=1.0\textwidth]{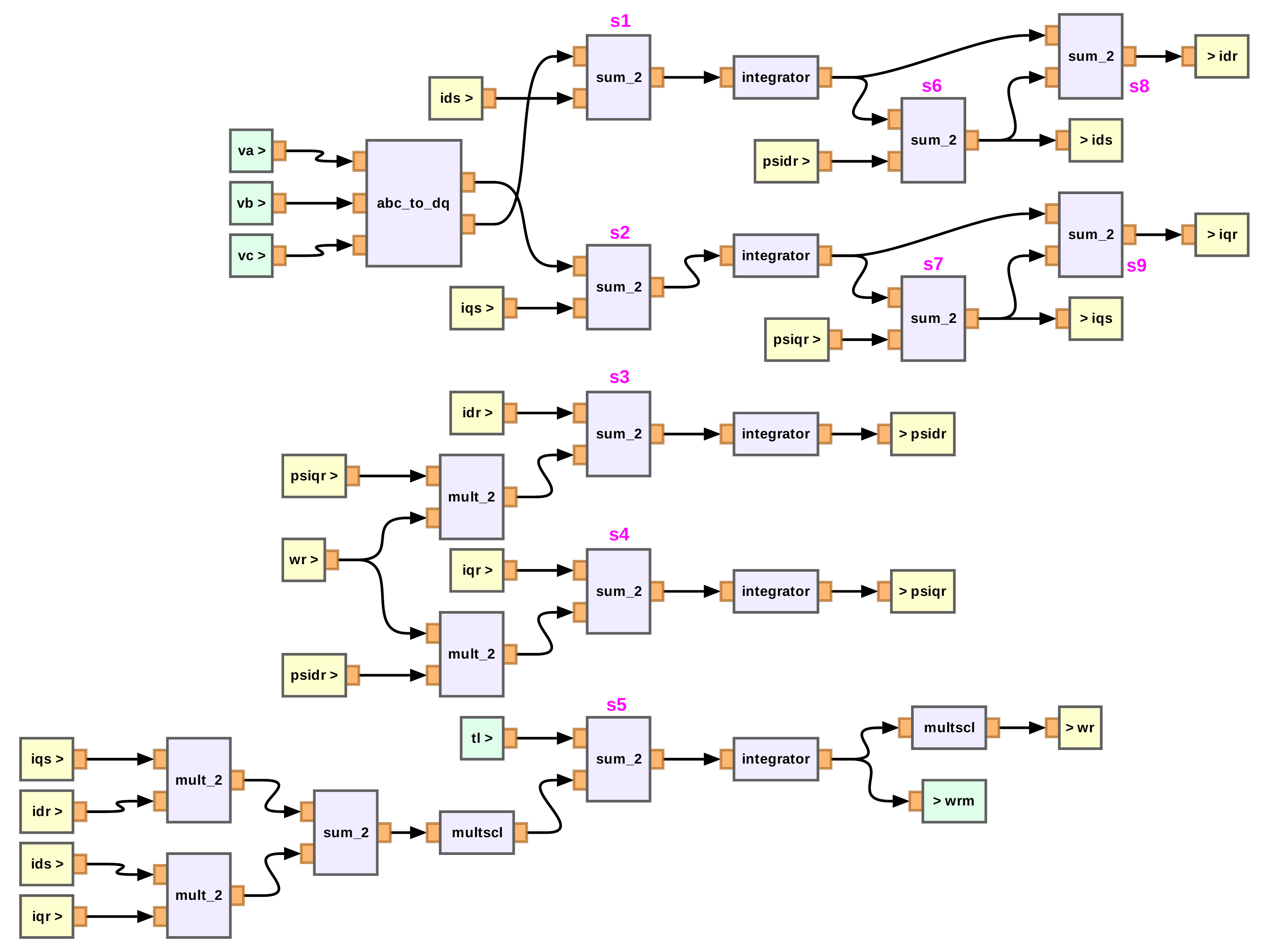}}
\vspace*{-0.2cm}
\caption{Subcircuit for the induction machine model given by
Eqs.~\ref{eq_indmc_1}-\ref{eq_indmc1_5}.}
\label{fig_sub_indmc}
\end{figure*}

As an example, Eq.~\ref{eq_indmc_3} can be rewritten as,
\begin{equation}
\psi _{dr} = \int \left(-\,\frac{P}{2}\,\omega _{rm}\psi _{qr}-r_r i_{dr}\right)\,dt,
\label{eq_indmc_3a}
\end{equation}
which can be implemented using a multiplier (to multiply $\omega _{rm}$ and
$\psi _{qr}$), and the {\tt{sum\_2}} and {\tt{integrator}} elements described
in Sec.~\ref{sec_templates}. Treating
Eqs.~\ref{eq_indmc_1}-\ref{eq_indmc_5}
in this manner, we obtain the subciruit shown in
Fig.~\ref{fig_sub_indmc}.

The following additional points about the implementation may be
noted:
\begin{list}{(\alph{cntr2})}{\usecounter{cntr2}}
 \item
  ``Virtual" sources and sinks (shown in light yellow colour) are used
  in order to make wiring less cumbersome. For example, note the virtual
  sink marked ``{\tt{>idr}}" and the virtual source marked
  ``{\tt{idr>}}", the two corresponding to the same node.
 \item
  Input and output pads (shown in light green colour) are used to indicate
  the input and output ports the subcircuit symbol will have when it is
  invoked (from a higher level). For the induction machine subcircuit,
  {\tt{vds}}, {\tt{vqs}}, {\tt{tl}} are the input ports, and
  {\tt{wrm}} is the output port.
 \item
  The subcircuit has the following parameters (not shown in Fig.~\ref{fig_sub_indmc}):
  {\tt{j}},
  {\tt{llr}},
  {\tt{lls}},
  {\tt{lm}},
  {\tt{poles}},
  {\tt{rr}},
  {\tt{rs}},
  which correspond to
  $J$,
  $L_{lr}$,
  $L_{ls}$,
  $L_m$,
  $P$,
  $r_r$,
  $r_s$,
  respectively, in 
  Eqs.~\ref{eq_indmc_1}-\ref{eq_indmc_5}.
  In implementing the equations, we need to compute quantities which
  depend on these parameters.
  For example, consider Eq.~\ref{eq_indmc1_1} for $i_{ds}$, implemented using
  the {\tt{sum\_2}} element marked as {\tt{s6}} in the figure. This element
  gives $i_{ds} \,$=$\, k_1\psi _{ds} + k_2\psi _{dr}$, which requires
  $k_1 \,$=$\, \displaystyle\frac{L_r}{L_mL_e}$,
  $k_2 \,$=$\, -\,\displaystyle\frac{1}{L_e}$ to be assigned.
  For all such assignments, the user is expected to supply a python function
  specific to the concerned subcircuit. For the induction machine subcircuit
  ({\tt{s\_indmc}}), the python block is shown in Fig.~\ref{fig_indmc_parm}.
  The calculations for $k_1$ and $k_2$ of {\tt{s6}} are shown specifically
  in the figure. Similarly, several other quantities are computed and stored
  in the python dictionary received by the {\tt{s\_indmc\_parm}} function
  as an argument.
  With this mechanism, the user has significant flexibility in implementing
  element equations.
\begin{figure}[!ht]
\centering
\scalebox{0.8}{\includegraphics{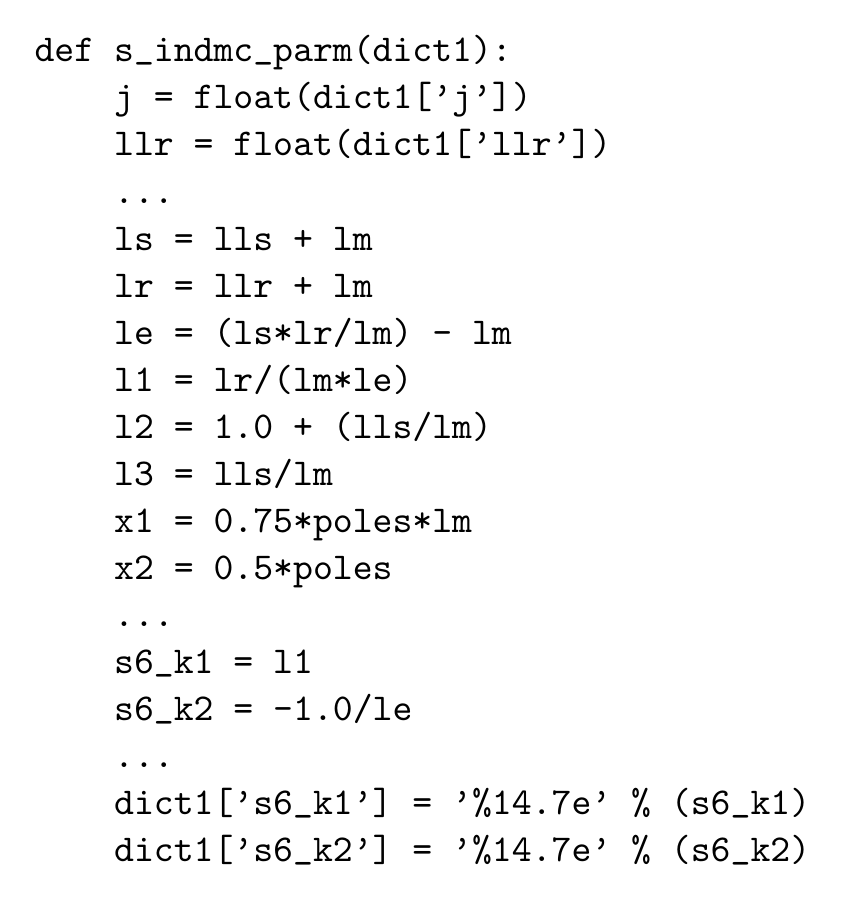}}
\vspace*{-0.2cm}
\caption{Parameter computation for the induction machine subcircuit.}
\label{fig_indmc_parm}
\end{figure}
 \item
  The user can define ``output parameters" for a subcircuit and use those at
  higher levels, as described in \cite{gseim_manual}. The output parameters
  can be mapped to nodes within the subcircuit or to the output parameters
  of the blocks involved in the subcircuit. These features provide a mechanism
  for viewing various quantities of interest at different levels when the
  system is simulated.
\end{list}

\section{Handling abrupt changes}
\label{sec_abrupt}
\begin{figure}[!ht]
\centering
\hspace*{-0.0cm}{\includegraphics[width=1.0\columnwidth]{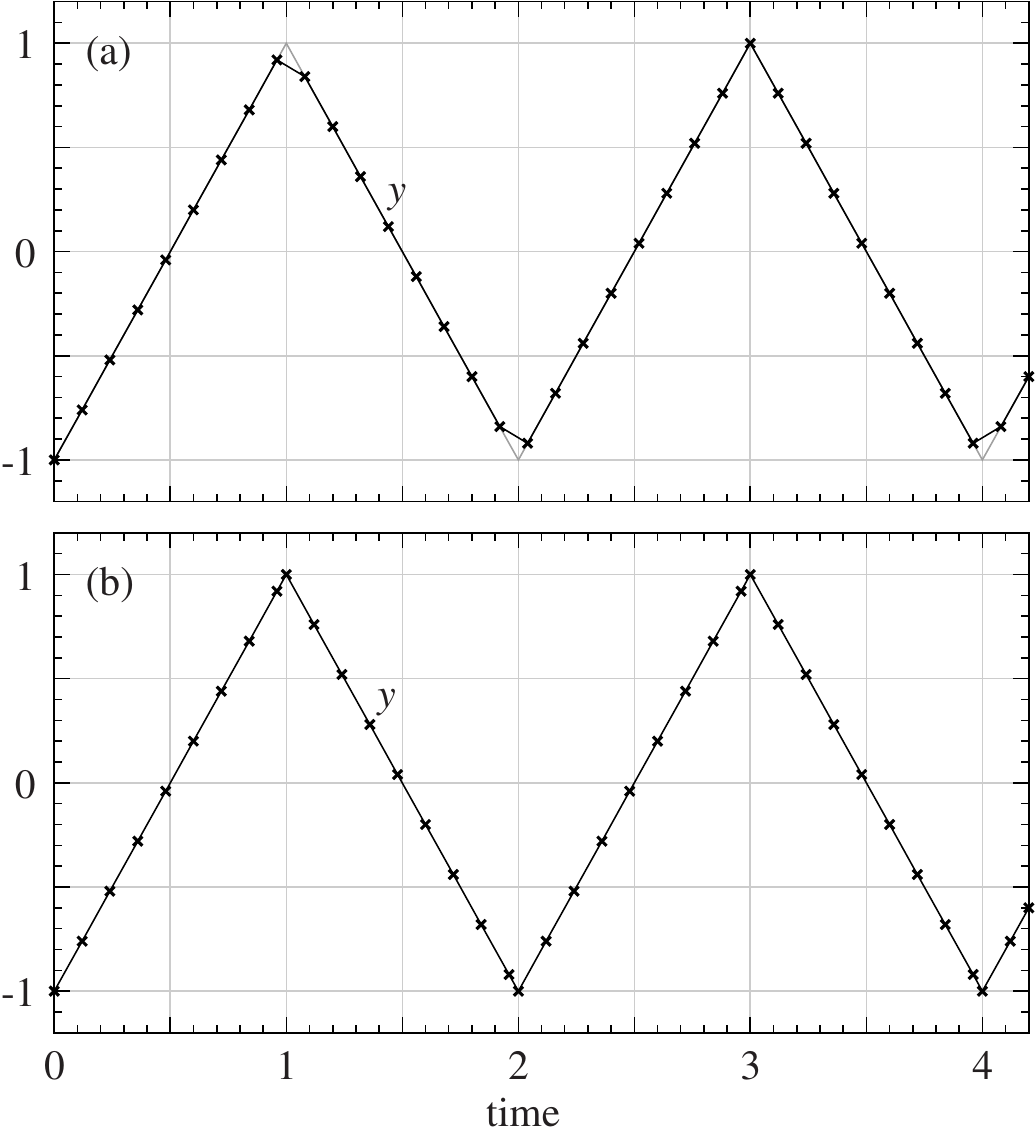}}
\vspace*{-0.2cm}
\caption{Triangle source waveform: (a)~with a constant
time step, (b)~with time step adjusted for tracking
abrupt changes. Crosses show the simulator time points.}
\label{fig_tri}
\end{figure}

In many systems of practical interest, some of the variables are
expected to vary abruptly. If the abrupt transitions are missed out
by the simulator, it would affect the appearance of the plots of
those variables, and more importantly, the accuracy of the simulation
results in some cases. As an example, consider a triangle source with
period $T \,$=$\, 2$\,s. If a constant time step $\Delta t \,$=$\, 0.12$\,s
is used, some of the peak or valley points are missed out by the simulator,
as seen in
Fig.~\ref{fig_tri}\,(a).
This situation can be improved by tracking the time points where the source
output is going to reach a peak or a valley. For this purpose, the triangle source template
in GSEIM takes the current time point from the simulator and returns the time of the
next ``break" (peak or valley). Using this information, GSEIM decides whether the normal
time step or a reduced time step should be used next time.
Fig.~\ref{fig_tri}\,(b) shows the triangle source waveform obtained with this approach.

\begin{figure}[!ht]
\centering
\hspace*{-0.0cm}{\includegraphics[width=1.0\columnwidth]{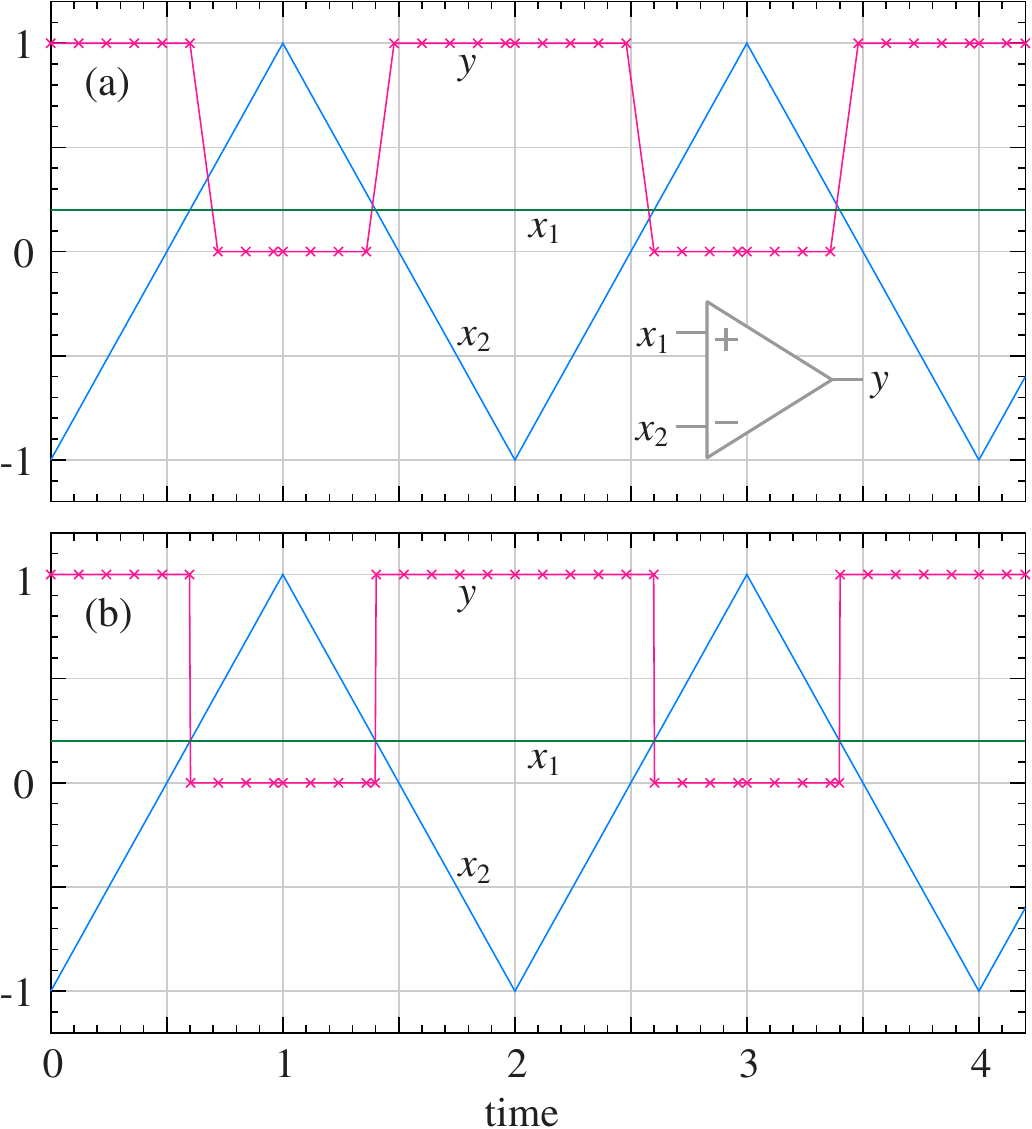}}
\vspace*{-0.2cm}
\caption{Comparator input and output waveforms: (a)~with a constant
time step, (b)~with additional time points obtained by extrapolation.
Crosses show the simulator time points.}
\label{fig_cmpr}
\end{figure}

Next, consider a comparator with inputs $x_1$, $x_2$ and output $y$.
Fig.~\ref{fig_cmpr}\,(a) shows the
$x_1$, $x_2$, and $y$ waveforms when a constant time step is used.
Abrupt changes in $x_2(t)$ are tracked by the simulator (as discussed
above); however, $y(t)$ is not resolved correctly by the simulator,
e.g., see the transition at $t \,$=$\, 2.6$\,s. One way to improve the
$y$ waveform is to uniformly reduce the time step, but this would make the
simulation slower, and from an accuracy perspective, small uniform time
steps may not even be required.

For resolving the abrupt transition in $y(t)$ on a shorter time scale
but without making the ``normal" time step (denoted by $\Delta t _{\mathrm{normal}}$)
small, GSEIM uses the following scheme. The comparator template stores the
previous time point $t_{-1}$ and the corresponding solutions
$x_1^{(-1)}$,
$x_2^{(-1)}$.
Knowing these, and the current time point and solutions
($t_0$, $x_1^{(0)}$, $x_2^{(0)}$), it uses linear extrapolation to compute
the time $t'$ at which $x_1-x_2$ would cross zero. If $t'$ is within
$\Delta t _{\mathrm{normal}}$ of $t_0$, GSEIM places one time point just before
and one time point just after $t'$.
Fig.~\ref{fig_cmpr}\,(b) shows the waveforms obtained with this method.
The transition at 
$t \,$=$\, 2.6$\,s is now seen to be resolved properly.

\begin{figure}[!ht]
\centering
\hspace*{-0.0cm}{\includegraphics[width=1.0\columnwidth]{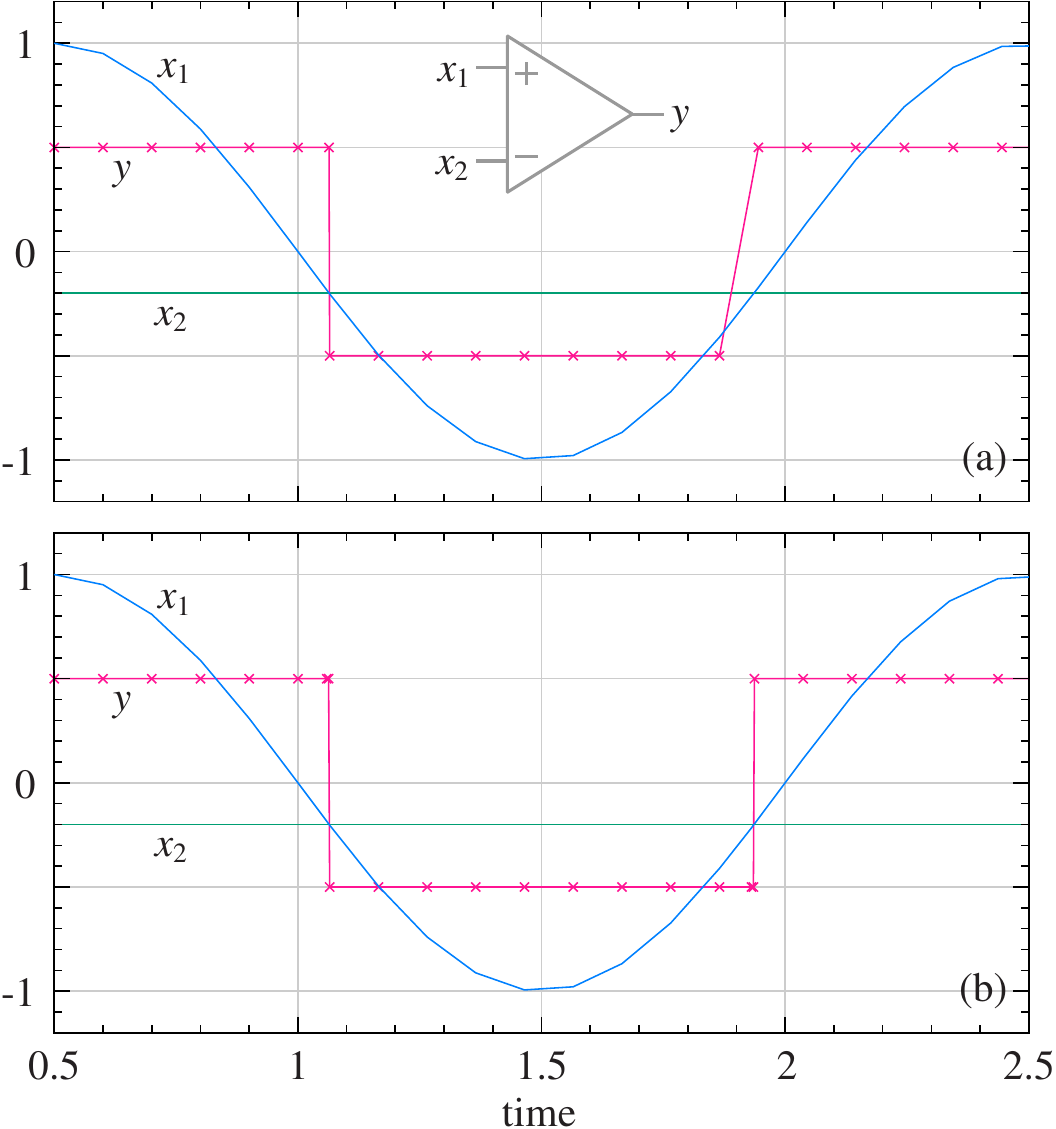}}
\vspace*{-0.2cm}
\caption{Comparator input and output waveforms:
(a)~with linear extrapolation,
(b)~with quadratic extrapolation.}
\label{fig_sin}
\end{figure}

\begin{figure}[!ht]
\centering
\scalebox{0.9}{\includegraphics{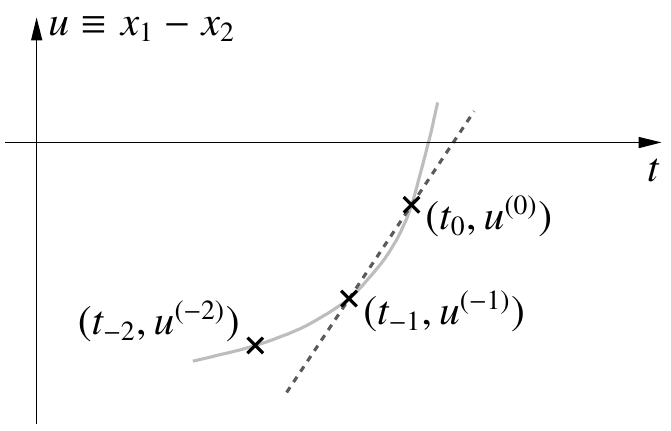}}
\vspace*{-0.2cm}
\caption{Example of failure of the linear extrapolation technique
for treating crossover events.}
\label{fig_quad}
\end{figure}

When the linear extrapolation technique is used for the comparator inputs shown in
Fig.~\ref{fig_sin}\,(a),
the zero crossing at $t \approx 1.93\,$s is not treated correctly because
linear extrapolation overestimates $t'$ in this case, as illustrated in
Fig.~\ref{fig_quad}. This problem can be addressed as follows. The comparator
stores, in addition to $t_0$, $t_{-1}$ (and the corresponding solutions), one
more past time point $t_{-2}$ and the corresponding solutions
$x_1^{(-2)}$,
$x_2^{(-2)}$.
It uses this information to fit a quadratic which passes through
$(t_0, u^{(0)})$,
$(t_{-1}, u^{(-1)})$,
$(t_{-2}, u^{(-2)})$,
(where $u\equiv x_1-x_2$), and computes $t'$ at which it goes through zero.
Fig.~\ref{fig_sin}\,(b) shows the waveforms obtained with the quadratic
extrapolation scheme. It can be seen that the transition
at $t \approx 1.93\,$s is now treated correctly.

\section{Simulation Examples}
\label{sec_examples}

We now look at simulation examples which demonstrate the capabilities
of GSEIM. We consider two examples, both involving the induction machine
model described by
Eqs.~\ref{eq_indmc_1}-\ref{eq_indmc_5}.
Details regarding setting up the schematic, running the simulation,
viewing the plots, will be described in \cite{gseim_manual} and are
not reproduced here. Also, a detailed description of the system being
simulated, values of the machine parameters, etc. are not included.
The simulation times mentioned in the following are for a desktop computer
(Ubuntu-19) with 3.2\,GHz clock and 8\,GB RAM.
\subsection{Free acceleration of induction motor}
\label{sec_free_indmc}
\begin{figure}[!ht]
\centering
\hspace*{-0.6cm}\scalebox{0.6}{\includegraphics{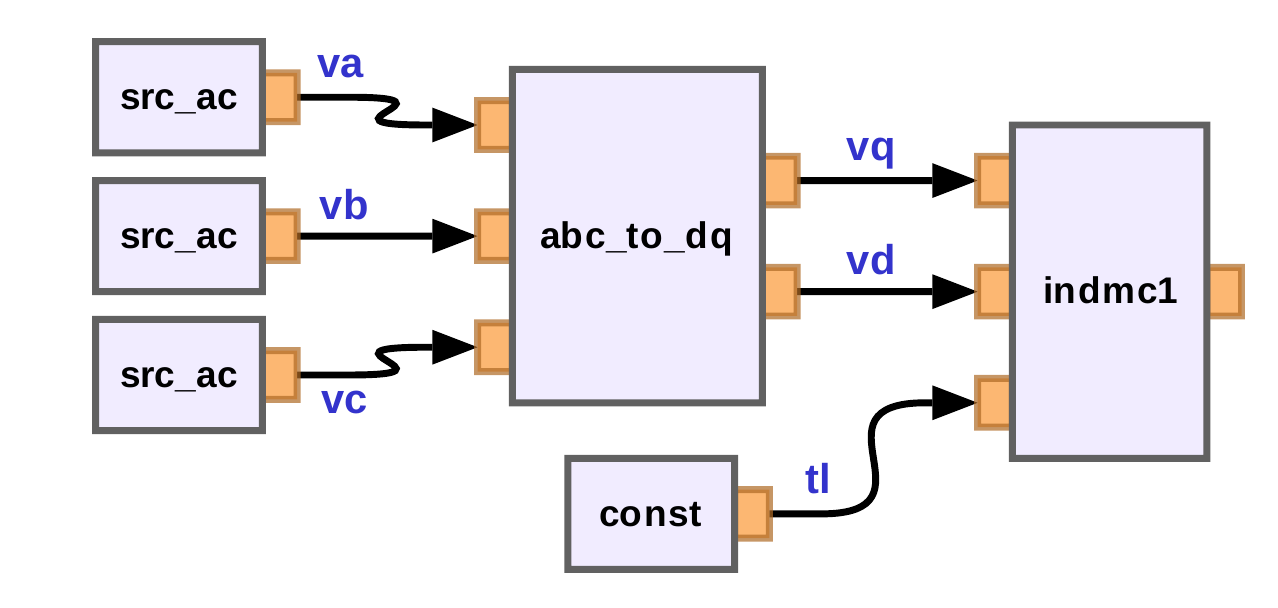}}
\vspace*{-0.2cm}
\caption{GSEIM schematic diagram for simulation of free acceleration
of induction motor using {\tt{indmc1.xbe}}}
\label{fig_freeacc_1}
\end{figure}

\begin{figure}[!ht]
\centering
\scalebox{0.6}{\includegraphics{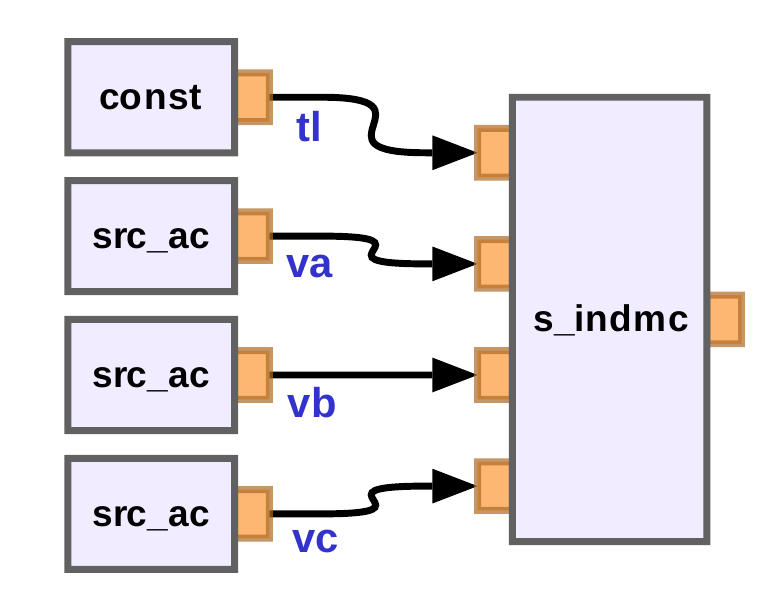}}
\vspace*{-0.2cm}
\caption{GSEIM schematic diagram for simulation of free acceleration
of induction motor using induction machine subcircuit.}
\label{fig_freeacc_2}
\end{figure}

\begin{figure}[!ht]
\centering
\hspace*{0.0cm}\scalebox{0.75}{\includegraphics{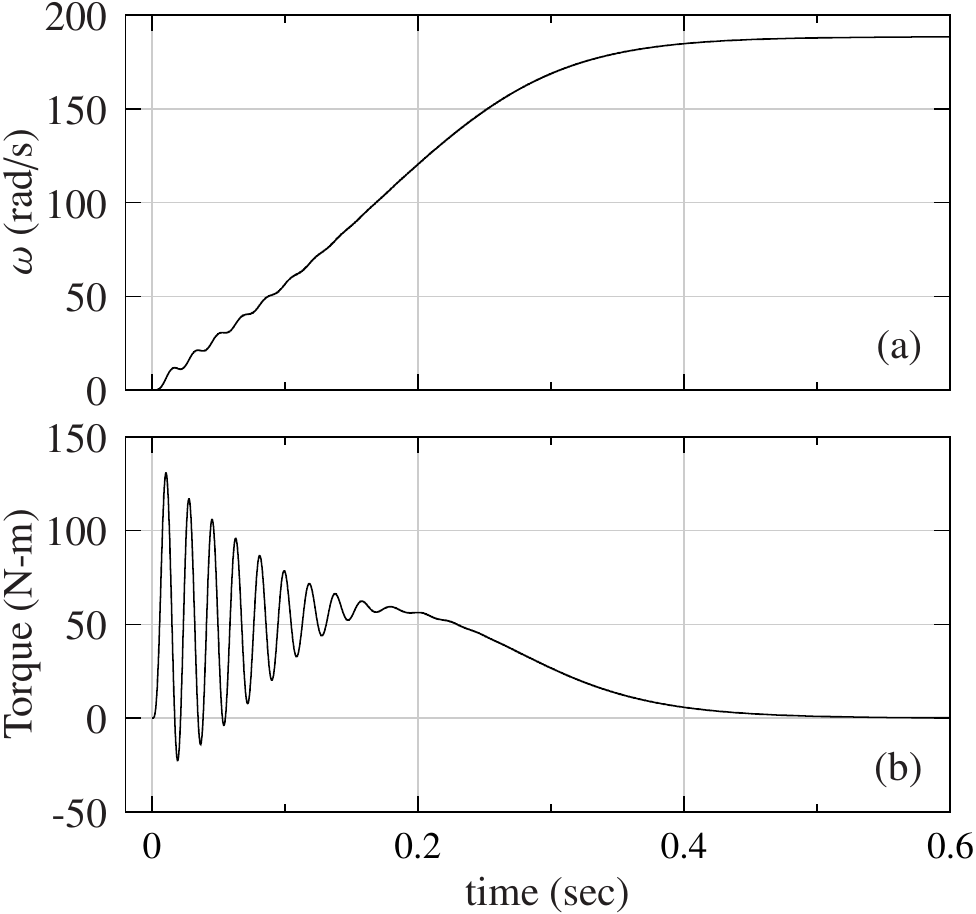}}
\vspace*{-0.2cm}
\caption{Simulation results for free acceleration
of induction motor: (a)~speed, (b)~torque}
\label{fig_freeacc_3}
\end{figure}

In this example, we consider free acceleration of an induction motor
with load torque $T_L \,$=$\, 0$.
Fig.~\ref{fig_freeacc_1} shows the GSEIM schematic diagram for the
system when the induction motor template discussed in Sec.~\ref{sec_templates}
is used. In this case, the conversion of
$V_a$,
$V_b$,
$V_c$ to
$V_d$,
$V_q$ is required, and it is performed by the {\tt{abc\_to\_dq}} element.
Fig.~\ref{fig_freeacc_2} shows the GSEIM schematic diagram for the
same problem when the induction machine subcircuit of Sec.~\ref{sec_subc}
is used. In this case, the $abc$-to-$dq$ conversion is incorporated
within the subcircuit.
Figs.~\ref{fig_freeacc_3}\,(a) and
\ref{fig_freeacc_3}\,(b)
show the simulation results for speed and torque, respectively.
The capability of GSEIM to produce plots of interest without
cluttering the schematic diagram with scopes may be noted.

\subsection{$V/f$ control of induction motor}
\label{sec_vbyf}
\begin{figure*}[!ht]
\centering
\hspace*{-0.1cm}{\includegraphics[width=1.0\textwidth]{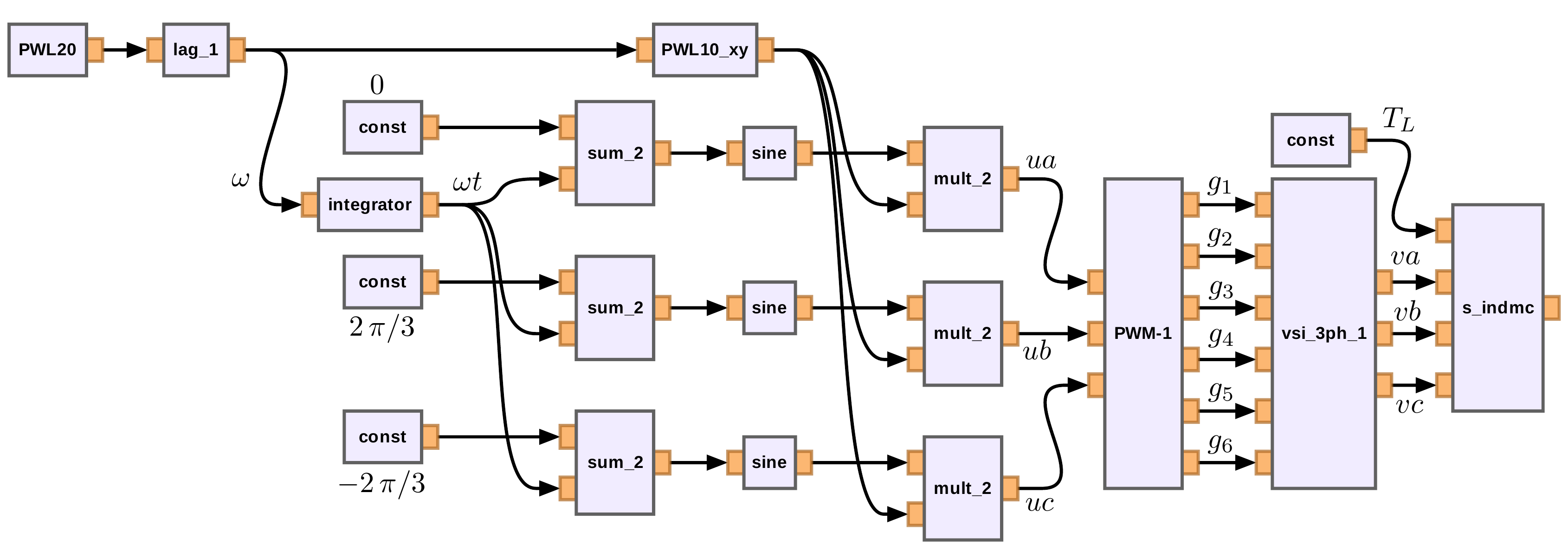}}
\vspace*{-0.2cm}
\caption{GSEIM schematic diagram for $V/f$ control of an induction motor.}
\label{fig_vbyf_1}
\end{figure*}

The GSEIM schematic diagram for $V/f$ control of an induction
motor\,\cite{mbpvrvtr} is shown in
Fig.~\ref{fig_vbyf_1}.
The induction machine block shown in the figure
corresponds to the subcircuit of
Fig.~\ref{fig_sub_indmc}; however, we can also use the element
template {\tt{indmc1.xbe}} (see Sec.~\ref{sec_templates}) directly.
The {\tt{pwl20}} element, which
allows piecewise linear waveforms, is used to generate the frequency command
which is smoothened using the {\tt{lag\_1}}
element, satisfying
\begin{equation}
\displaystyle\frac{dy}{dt} =\displaystyle\frac{1}{T_r}\,(-y+x).
\label{eq_lag}
\end{equation}
\noindent
The $V/f$ conversion is provided by {\tt{pwl10\_xy}}.

\begin{figure*}[!ht]
\centering
\scalebox{0.6}{\includegraphics{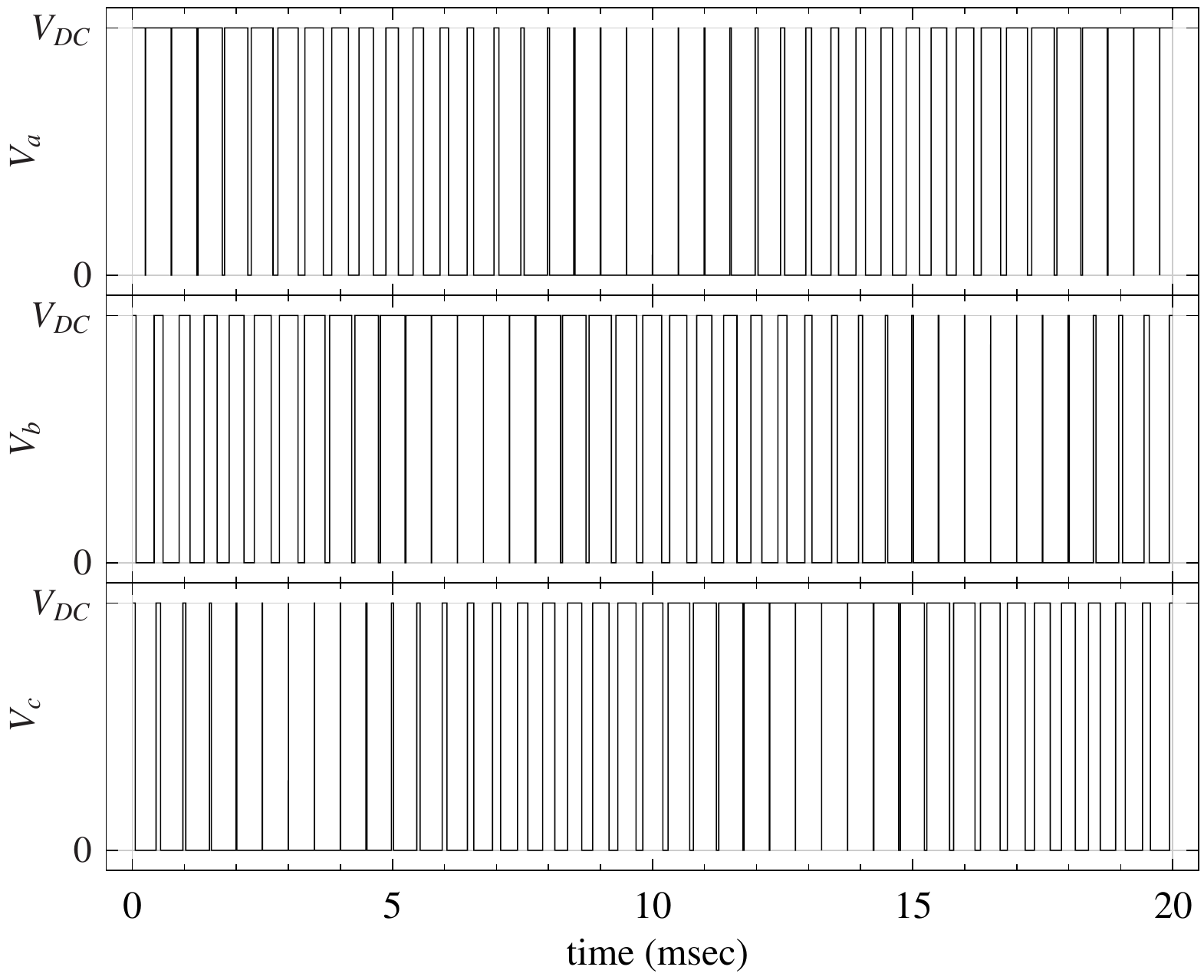}}
\vspace*{-0.2cm}
\caption{Voltages $V_a$, $V_b$, $V_c$ in steady state for the system
of Fig.~\ref{fig_vbyf_1}.}
\label{fig_vbyf_2}
\end{figure*}

\begin{figure}[!ht]
\centering
\scalebox{0.75}{\includegraphics{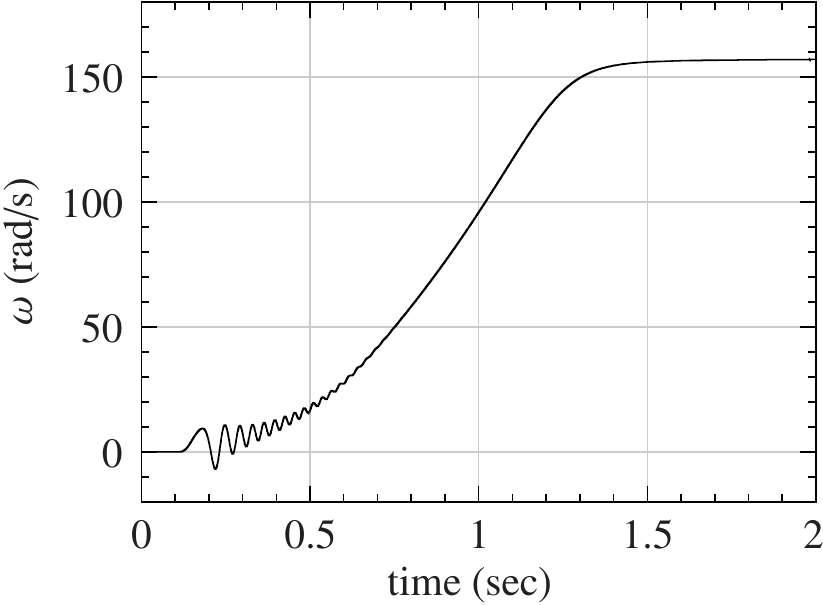}}
\vspace*{-0.2cm}
\caption{Speed versus time for the system of Fig.~\ref{fig_vbyf_1}.}
\label{fig_vbyf_3}
\end{figure}

Simulation of this system is computationally more demanding than
the free acceleration example because (a)~its size is larger,
(b)~accurate resolution of PWM voltages requires smaller time steps.
For resolving the PWM waveforms correctly, the linear extrapolation
technique described in Sec.~\ref{sec_abrupt} has been used.
Fig.~\ref{fig_vbyf_2} shows the PWM voltages in steady state.
It is seen that the transitions between low and high levels are
treated properly.
Fig.~\ref{fig_vbyf_3} shows the motor speed versus time.

It is interesting to compare the simulation times when two different
approaches are used: (a)~induction machine as a subcircuit (described
in Sec.~\ref{sec_subc}), (b)~induction machine as an element template
(described in Sec.~\ref{sec_templates}). To enable a fair comparison,
the same numerical method (RKF45) was used in both cases, and the algorithmic
parameters were kept the same. The simulation times were found to be
2.33\,s and 1.55\,s, respectively, for (a) and (b). This clearly brings out
the advantage of implementing equations at a low level (the template level)
rather than as a subcircuit. On the other hand, subcircuit implementation is
often easier, and therefore it may be preferred when simulation time is not
large enough to be of concern.

\section{Conclusions and future work}
\label{sec_conclusions}
In summary, we have presented a new general-purpose ODE solver called
GSEIM which allows the use of explicit as well as implicit methods.
The organisation of the program has been described. A useful feature of
GSEIM is that it allows the user to write new templates to incorporate
element equations. In addition, it also allows the use of subcircuits
(hierarchical blocks). These two facilities are illustrated with the help of examples.
The importance of correct handling of abrupt changes is pointed out, and
the techniques used by GSEIM to handle abrupt changes are explained.
Finally, two representative simulation examples are presented. The computational
advantage of incorporating element equations in a basic template rather than
a subcircuit is brought out.

The authors plan to make the GSEIM package available under the GNU
general public license, after preparing a users' manual and video
tutorials. Further developments in GSEIM are expected to address the
following issues.
\begin{list}{(\alph{cntr2})}{\usecounter{cntr2}}
 \item
  Additional features in the schematic capture GUI such as rectangular
  wires, ``bus" facility (i.e., grouping of several connections into a
  single wire), use of images in the element symbols, provision for
  adding text to the canvas, etc.
 \item
  Allowing electrical type elements, which satisfy some form of Kirchhoff's
  current and voltage laws, in addition to the ``flow-graph" type elements
  currently available. This will significantly enhance the applications
  handled by GSEIM.
\end{list}

\bibliographystyle{IEEEtran}
\bibliography{ref1}

\end{document}